\documentclass[a4paper,fleqn]{cas-dc}

\usepackage[numbers]{natbib}
\usepackage{amsmath,bm}
\newcommand{\SUN}{S.U.N.}
\usepackage{graphicx}
\usepackage{xcolor}
\usepackage{tikz}
\usetikzlibrary{shapes.geometric, arrows.meta, positioning, fit, backgrounds}
\usepackage{float}
\floatstyle{ruled}
\newfloat{algorithm}{tbp}{loa}
\floatname{algorithm}{Algorithm}
\newcounter{algline}
\newcommand{\kw}[1]{\textbf{#1}}
\newcommand{\proc}[1]{\textsc{#1}}
\newcommand{\cmt}[1]{\,$\triangleright$~\textit{#1}}
\newcommand{\ind}{\hspace*{1.4em}}
\newcommand{\indd}{\hspace*{2.8em}}
\newenvironment{pseudo}
 {\footnotesize\begin{list}{\arabic{algline}:}%
   {\usecounter{algline}%
    \setlength{\leftmargin}{2.4em}\setlength{\labelwidth}{1.8em}%
    \setlength{\labelsep}{0.4em}\setlength{\itemsep}{1.5pt}%
    \setlength{\parsep}{0pt}\setlength{\topsep}{3pt}\setlength{\partopsep}{0pt}}}
 {\end{list}}

\begin{document}
\let\WriteBookmarks\relax
\def\floatpagepagefraction{1}
\def\textpagefraction{.001}

\shorttitle{MatEvolve: crystal-structure design in a language of motifs}
\shortauthors{D.-K. Le et~al.}

\title[mode = title]{Crystal-structure design by agentic AI in a language of motifs}

\author[1]{Dinh-Khiet Le}

\author[1]{Minh-Quyet Ha}[orcid=0000-0003-4617-0059]
\cormark[1]
\ead{mq-ha@jaist.ac.jp}

\author[1]{Hong-Phuc Vu-Dinh}

\author[2]{Viet-Cuong Nguyen}[orcid=0000-0002-8008-582X]

\author[3]{Huan Tran}[orcid=0000-0002-8093-9426]

\author[4]{Truyen Tran}[orcid=0000-0001-6531-8907]

\author[5]{Takashi Miyake}[orcid=0000-0003-2658-3470]

\author[6]{Hiori Kino}[orcid=0000-0002-8912-686X]

\author[1,7]{Hieu-Chi Dam}[orcid=0000-0001-8252-7719]
\cormark[1]
\ead{dam@jaist.ac.jp}

\affiliation[1]{organization={Japan Advanced Institute of Science and Technology},
                addressline={1-1 Asahidai}, 
                city={Nomi},
                state={Ishikawa 923-1292},
                country={Japan}}

\affiliation[2]{organization={HPC SYSTEMS Inc.},
                addressline={3-9-15 Kaigan},
                city={Minato-ku},
                state={Tokyo 108-0022},
                country={Japan}}     
                
\affiliation[3]{organization={School of Materials Science and Engineering, Georgia Institute of Technology},
                addressline={771 Ferst Dr. NW},
                city={Atlanta},
                state={Georgia 30332},
                country={United States}}

\affiliation[4]{organization={Applied Artificial Intelligence Institute, Deakin University},
                city={Geelong},
                state={Victoria 3216},
                country={Australia}}

\affiliation[5]{organization={Department of Physics, School of Science, Institute of Science Tokyo},
                addressline={2-12-1 Ookayama},
                city={Meguro-ku},
                state={Tokyo 152-8551},
                country={Japan}}

\affiliation[6]{organization={Research Center for Materials Informatics, Department of Advanced Data Science, The Institute of Statistical Mathematics},
                addressline={10-3 Midori-cho}, 
                city={Tachikawa},
                state={Tokyo 190-8562},
                country={Japan}}
                
\affiliation[7]{organization={International Center for Synchrotron Radiation Innovation Smart (SRIS), Tohoku University},
                addressline={2-1-1 Katahira}, 
                city={Aoba-ku},
                state={Sendai 980-8577}, 
                country={Japan}}

\cortext[cor1]{Corresponding author}

\begin{abstract}
Data-driven materials discovery interpolates more reliably than it extrapolates and seldom reveals new structure types. We present MatEvolve, an agentic-AI framework that designs crystals in an interpretable \emph{language of motifs}, in which each crystal is written as a \emph{motif profile} that describes the recurring geometric patterns---the \emph{motifs}---composing the crystal. The motif profile serves not merely as a description of a material but as the medium for material design: the agent edits the profile with a stated rationale and constructs a crystal from the modified one; the most promising candidates are then validated by first-principles calculations. MatEvolve is built on the state-of-the-art language model Claude Fable~5 without fine-tuning. It is applied to the design of rare-earth-lean permanent-magnet compounds and finds novel (meta)stable crystals that are both new structural prototypes and magnet compounds; under an equal validation budget, it does so an order of magnitude more often than generative models. Beyond design, analyzing the human-readable profiles of the discovered crystals reveals their structure--property relationships.
\end{abstract}


\maketitle


\section{Introduction}

\begin{figure*}[t]
    \centering
    \includegraphics[width=0.95\textwidth]{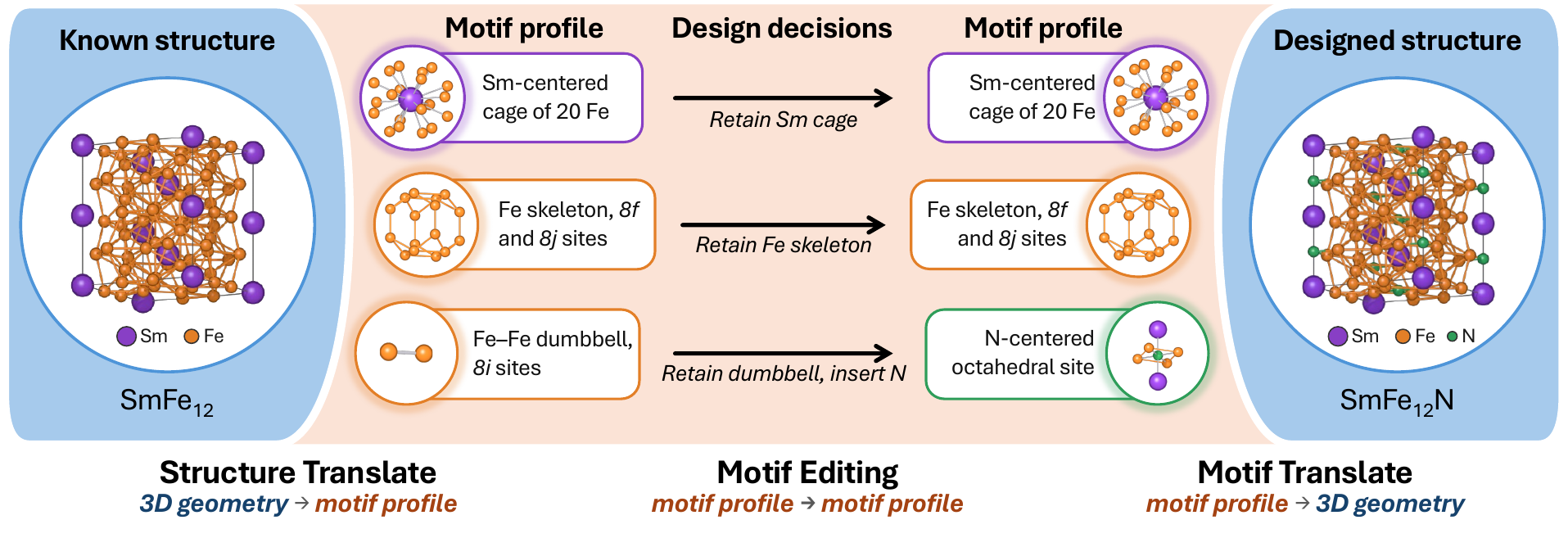}
\caption{\textbf{The motif profile as the medium of crystal design.} One design step, shown for ThMn$_{12}$-type SmFe$_{12}$ and its interstitial nitride SmFe$_{12}$N. Left: Structure Translate, an agent drawing on crystallographic knowledge, reads the whole known crystal and writes its motif profile, a text that names the motifs from which the crystal is built---here a Sm-centered cage of 20 Fe atoms, Fe--Fe dumbbells, and the Fe skeleton that links them. The images beside the motifs illustrate what the text names; the crystal is not cut into fragments. Center: Motif editing edits the profile; each decision (retain, insert) carries a rationale, in quotation marks, drawn from domain knowledge. Right: Motif Translate builds the whole three-dimensional structure from the edited profile.}
    \label{fig:workflow}
\end{figure*}

New materials have repeatedly redefined the limits of technology, from semiconductors underpinning modern computing to magnets and electrodes used in clean-energy systems. However, their discovery remains slow and costly: the space of candidate structures is effectively infinite, each candidate is expensive to synthesize and characterize. Moreover, after a promising material is identified, translating it into practical applications can take a decade or more. To accelerate discovery, materials science has progressed through four established paradigms---empirical experiment, theory, simulation, and data-driven learning~\cite{hey2009}---and the fifth is now emerging, in which AI agents drive discovery~\cite{berens2023,bishop2022ai4science,ioannidis2024fifthparadigm}. Setting the empirical baseline aside, we consider the computational paradigms in turn---theory and simulation, data-driven learning, and the emerging agent-based approach---asking what each contributes, where it falls short, and how its limitations motivate the next paradigm.

First-principles theory and simulation reason deductively; they predict how a given structure will behave from the established physical theory. Density-functional theory (DFT) does so through a tractable approximation, computing the stability and properties of a structure by calculation alone~\cite{Jain2013}. Its distinctive strength is that it can assess a material that does not yet exist both quantitatively and with an accuracy determined by the underlying approximation, thereby helping determine whether it is worth synthesizing and characterizing. However, it faces two limitations as a route to discovery. First, each calculation is expensive; thus, even large-scale parallel computation covers only a negligible fraction of an effectively infinite candidate space. Second, deduction evaluates candidates but does not generate them; that task falls to other means: enumeration by researchers, crystal-chemistry rules such as Pauling's rules~\cite{pauling1929}, or structure-search algorithms such as the evolutionary search~\cite{oganov2006uspex} and particle-swarm search~\cite{wang2012calypso}. However, these approaches are also limited: enumeration and rule-based generation largely recombine known chemistry, while searching remains constrained by the high computational cost, which can make calculations infeasible for complex structures. These limitations have motivated an inductive, data-driven paradigm.

The data-driven paradigm learns statistical structure--property relationships from large, curated datasets, reasoning by induction---generalizing from many examples~\cite{Dam2026}. These inductive methods take two forms. \emph{Surrogates} predict stability and property orders of magnitude faster than DFT at inference, allowing far larger candidate sets to be screened~\cite{choudhary2021alignn,chen2019megnet,deng2023chgnet,chen2022m3gnet,Vu2023}. \emph{Generators} learn a distribution over known structures and sample new candidates to supply that screen; some are conditioned on a target property~\cite{xie2022cdvae,jiao2023diffcsp,mattergen2025,miller2024flowmm,gruver2024llm_materials,antunes2024crystallm}. The largest such studies have predicted large numbers of DFT-stable candidates~\cite{merchant2023gnome}. Active learning and autonomous laboratories add a selection layer, ranking candidates using surrogate estimates such as predicted stability or uncertainty to determine which candidates to compute or synthesize next, and can automate the experiments themselves~\cite{Lookman2019,szymanski2023alab}.

These inductive methods are most reliable where they interpolate: within the range of known materials they predict well, but their accuracy tends to decrease beyond it~\cite{meredig2018extrapolation}---and that is where genuinely new materials lie. A secondary trade-off concerns interpretability: the most accurate surrogates are often the least transparent, whereas interpretable descriptors such as symbolic regression~\cite{ouyang2018sisso} can recover physical insight at some cost in accuracy. The catch is what counts as \emph{new}. A generator samples from a distribution fitted to known structures; therefore, a compound merely absent from current databases is readily obtained. A genuinely new \emph{prototype}---an arrangement unlike those in the training set---tends to lie in the low-probability tail of that distribution and is reached only rarely~\cite{cheetham2024ai}. Moreover, a likely draw provides little rationale for why it should meet the target property. Such methods generate readily within the known space but less often reach a new, better prototype---a gap that the next paradigm may help close.

The emerging agent-based approach addresses that task. At its center is an \emph{agent}---a large language model directed not by a task-specific dataset but by a goal stated in words~\cite{berens2023,wang2023aiscience}. It reasons abductively, drawing on the human knowledge absorbed during training to propose a plausible candidate and reason for it, which a subsequent test then confirms or refutes. Such agents already operate at research scale: language-model agents have autonomously identified real software vulnerabilities, agent-proposed scientific hypotheses have been confirmed in the laboratory~\cite{anthropic2026fablemythos,natarajan2026coscientist}, and agent systems such as ChatMOF answer materials queries and generate metal--organic frameworks by coordinating search, prediction, and generation tools~\cite{kang2024chatmof}. This approach is widely argued to mark an emerging fifth paradigm~\cite{bishop2022ai4science,ioannidis2024fifthparadigm}; the agent is itself trained on data, but a single agent can serve many goals, whereas the data-driven paradigm fits a new model to each task. Because the agent verbalizes its reasoning, it and the scientist share a medium through which the exchange can run in both directions: a scientist can shape the agent's proposals, and the agent makes its reasoning transparent in return. This reciprocity is something that one-directional couplings of human knowledge to a model, such as physics-informed learning~\cite{raissi2019pinn}, cannot readily offer.

In materials design, the goal is to identify a compound that is both stable and exhibits a target property. At a fixed composition, both are governed largely by how its atoms are arranged---the crystal's geometry. Geometry is something an agent can handle in words: a language model can solve an Olympiad geometry problem from its written statement, proposing the constructions required by its proof~\cite{trinh2024alphageometry}. A crystal can be translated into the same medium~\cite{le2026crystallography}: as the set of local motifs from which it is built, its \emph{motif profile}. From this text, an agent proposes edits that recombine and mutate parent motifs, as in an evolutionary search, but guided by a stated, target-directed reason rather than the fixed, target-blind operators of structure-search methods such as USPEX~\cite{oganov2006uspex} and CALYPSO~\cite{wang2012calypso}. The agent draws on knowledge far broader than any property-labeled dataset; therefore, its edits are not confined to the structures in that dataset, whereas data-driven generators mostly produce structures close to those they were trained on~\cite{Ha2025}.

Prior work has established each side of this crystal--text translation. Automated description turns a crystal into text: Robocrystallographer writes a structure as prose that reports its space group and prototype~\cite{ganose2019robocrystallographer}. Text-conditioned generation turns a specification back into a crystal: CrysText produces a structure from its composition and symmetry~\cite{mohanty2026crystext}. Textual descriptions further act as machine-readable representations for property prediction~\cite{rubungo2025llmprop,munjal2024latticelingo}, with the choice of representation itself shaping model performance~\cite{ozawa2026scaledependent}. Agentic loops have been demonstrated as well: FunSearch evolves computer programs~\cite{romera2024funsearch}, LLMatDesign and MatAgent iterate over chemical compositions~\cite{jia2024llmatdesign,takahara2025matagent}, and MatLLMSearch evolves raw crystal encodings~\cite{gan2025matllmsearch}. The closest of these, MatAgent, has a language model reason about which composition to try but leaves the atomic arrangement to a diffusion model trained on known crystals~\cite{takahara2025matagent}. Together, these studies provide every component: a crystal can be written as text, text can be rebuilt into a crystal, and an agent can steer an iterative search. They point to a natural next step---a framework in which text itself becomes the medium of materials design.

Herein, we present MatEvolve, a framework whose design medium is a \emph{language of motifs}: a crystal is described as a \emph{motif profile}, an agent edits the profile, and a new crystal is built from the edited profile (Fig.~\ref{fig:workflow}). The profile is a human-readable text that records the recurring geometric patterns---the \emph{motifs}---from which the crystal is built. This describe--edit--build cycle repeats as an evolutionary search: parent profiles are recombined or mutated with a stated rationale, an evidence-based recommender decides where the limited first-principles budget is spent, and each validated candidate joins the pool from which the parents of the next generation are drawn. A profile names the building blocks themselves---withholding the coordinates, lattice parameters, space group, and structure-type label that a crystallographic description would report. However, it is accepted only if a crystal containing its motifs can be built and, when the profile is translated from a known crystal, only if that crystal is recovered in a round-trip test. Thus, the agent edits the architecture of a crystal, not its coordinates.


\section{Results}
\label{sec:results}

\subsection{Framework overview}

\begin{figure*}[t]
    \centering
    \includegraphics[width=0.95\textwidth]{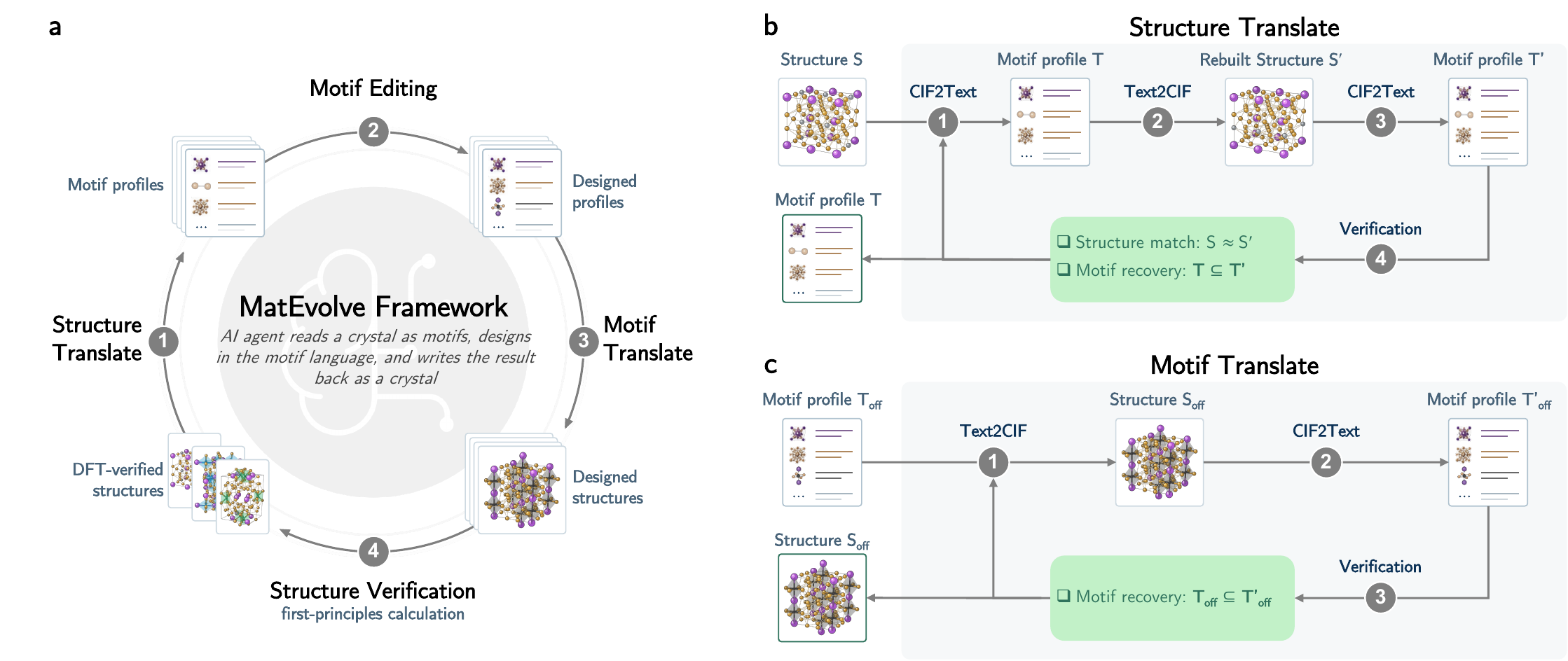}
\caption{\label{fig:workflow_diagram}\textbf{The MatEvolve loop and its two translation steps.} (a) One generation. Parent structures from the pool are written as motif profiles (1, Structure Translate), edited by the design agent with a stated rationale (2, Motif editing), realized as three-dimensional structures (3, Motif Translate), and validated by first-principles calculation (4, Structure Verification); validated structures join the pool from which the parents of the next generation are drawn. The evidence-based recommender that ranks candidates before validation is not drawn. (b) Structure Translate of a reference crystal $S$: CIF2Text writes its profile $T$ (1), Text2CIF rebuilds $S'$ from $T$ alone (2), and CIF2Text reads $S'$ back as $T'$ (3). $T$ is accepted if $S'$ matches $S$ (StructureMatcher) and every motif of $T$ is recovered in $T'$ (4); if either check fails, $T$ is redrafted, for up to five attempts. (c) Motif Translate of an offspring profile $T_{\mathrm{off}}$: Text2CIF builds a candidate structure $S_{\mathrm{off}}$ (1), and CIF2Text reads it back as $T'_{\mathrm{off}}$ (2). $S_{\mathrm{off}}$ is accepted if every motif of $T_{\mathrm{off}}$ is recovered in $T'_{\mathrm{off}}$ (3); otherwise it is rebuilt, for up to five attempts.}
\end{figure*}

MatEvolve is an agent-based framework for materials \emph{design} through evolutionary search. Rather than screening a fixed or sampled candidate pool, it evolves new crystals over an interpretable text representation---the \emph{motif profile}---across successive generations and validates its designs from first principles. It is designed to be retargetable, with a new objective set by restating it rather than by re-engineering the pipeline. Each generation runs as an ordered pipeline: \emph{Structure Translate} writes each parent crystal as its motif profile; \emph{Motif editing} edits those parent profiles into offspring profiles; \emph{Motif Translate} realizes each offspring profile as a crystal; an evidence-based recommender system (ERS) ranks the offspring; and DFT validates the top-ranked candidates (\emph{Structure Verification}), which then join the population from which the parents of the next generation are drawn. Structure Translate and Motif Translate share two basis agents that translate between a crystal and text in opposite directions---CIF2Text, which reads a structure and writes its profile, and Text2CIF, which generates a structure from a profile. The remainder of this section describes the design of each stage; implementation details are deferred to the Methods.

The motif profile itself is at the core of the framework. A crystal is written as a set of distinct structural motifs from which it is built, together with its composition and no atomic coordinates, lattice parameters, space-group symbol, or structure-type label. A motif is a recurring structural building block, ranging from a local coordination environment---a polyhedron, cage, or bonded cluster---to an extended framework such as a net or layer, each named in a controlled natural-language vocabulary according to its constituent species, geometry, and bond-length regime (Supplementary Information~S1). For example, a rare-earth--iron compound may be profiled as a samarium-centered high-coordination (Frank--Kasper-type) cage, an extended iron framework with short Fe--Fe contacts, and an interstitial light-element site. By naming the building blocks themselves rather than encoding coordinates, the profile is human-interpretable and directly editable---a materials scientist can read, edit, and compare it motif by motif---unlike a learned latent vector or a raw coordinate list. A profile must retain sufficient information to specify the crystal to be admissible: we require that the structure can be regenerated from it, satisfying a \emph{bidirectional-consistency} criterion enforced by the process described next.

\emph{Structure Translate} turns a crystal into an admissible motif profile, and this is where bidirectional consistency---the round-trip test---is enforced (Fig.~\ref{fig:workflow_diagram}b). Given a structure $S$, CIF2Text writes a candidate profile $T$; Text2CIF then reconstructs a structure $S'$ from $T$ alone---the same profile-to-crystal step that Motif Translate performs---and CIF2Text reads $S'$ back as a profile $T'$. $T$ is accepted only if $S'$ matches $S$ and every motif of $T$ is recovered in $T'$, by the criteria described in Methods~2. Because Text2CIF works from the profile, the agreement reflects the information carried by $T$; a profile from which no matching structure can be recovered is revised and, if necessary, rejected. Admissibility requires only that a faithful realization be recoverable, not that the profile uniquely determines a structure. CIF2Text and Text2CIF extend the text representation of crystals that we introduced previously~\cite{le2026crystallography}; their construction is given in Supplementary Information~S1 and Supplementary Information~S2, and the accept-or-revise loop is described in Methods~2.

\emph{Motif editing} turns parent profiles into an offspring profile (Fig.~\ref{fig:workflow_diagram}a, step 2). A design agent reads the profiles of one or two parents---drawn from the pool according to the design objective---and reasoning over their motifs, proposes an offspring profile $T_{\mathrm{off}}$ by recombining or mutating them (an evolutionary operator; Supplementary Information~S3) to improve a target property. \emph{Motif Translate} then realizes the offspring profile as a crystal (Fig.~\ref{fig:workflow_diagram}c): Text2CIF generates a candidate structure $S_{\mathrm{off}}$ from $T_{\mathrm{off}}$ (Methods~3), CIF2Text reads it back as $T'_{\mathrm{off}}$, and $S_{\mathrm{off}}$ is regenerated, for up to five attempts, until every designed motif is recovered, so that the design is faithfully realized.

Each generation proposes more candidates than can be afforded by first-principles validation; therefore, the ERS~\cite{Ha2021,Ha2025} ranks them beforehand. It scores a candidate not from its individual motifs but from its \emph{motif groups}---families of related motifs obtained by clustering them across the dataset (Methods~4). Through Dempster--Shafer evidence theory~\cite{dempster1967,shafer1976}, it accumulates evidence of how reliably those groups and their combinations have been associated with the design target across the labeled structures. Candidates are ranked by this evidence so that costly validation is focused on the most promising ones, and DFT then relaxes and evaluates each selected candidate. The evidence model and the first-principles validation protocol are described in Methods~5 and Methods~6.

\subsection{Experimental design and settings}
\label{sec:experiments}

We evaluated MatEvolve in two experiments. Experiment~1 investigates whether a crystal can be written as a motif profile and rebuilt from that profile alone. Experiment~2 verifies whether editing profiles yields stable, novel crystals that are magnet compounds and new prototypes at rates that are comparable to generative models under an equal validation budget, and how much of that performance comes from the recommender that selects candidates for validation. A descriptive analysis of the motif content of the resulting structures, prompted by the outcome of Experiment~2, follows in Section~2.5. All agents ran on a single frozen commercial language model (Claude Fable~5, API identifier \texttt{claude-fable-5}; decoding temperature $0.5$) that was identical across the experiments.

\paragraph{Datasets} Two datasets underpinned the experiments: one seeded the search, the other defined novelty.
\begin{itemize}
\item \textbf{Initial population, $\mathcal{D}_{\mathrm{seed}}$} --- the 948 magnetic structures from which the search started were curated from the Materials Project~\cite{Jain2013} to cover permanent-magnet and magnetic-intermetallic families: compounds pairing a rare-earth element (Ce, Pr, Nd, Sm, Gd, Tb, Dy, Ho, Er, Y) with a transition metal (Fe, Co, Mn, Ni), with total magnetization above $10~\mu_B$ per formula unit, energy above the convex hull of at most $0.2$~eV/atom, and converged DFT energies (tolerance below $10^{-5}$~eV/atom). Of these, 208 are thermodynamically stable ($E_{\text{hull}}=0$); by magnetic ordering, it included 282 ferromagnetic and 616 ferrimagnetic structures, with the remaining 50 unresolved in the source data. Compositions ranged from binary to quinary with 5--100 atoms per cell, including ThMn$_{12}$-type intermetallics, RCo$_5$ compounds, R$_2$T$_{14}$B compounds, Heusler alloys, and perovskite oxides.
\item \textbf{Novelty reference, $\mathcal{D}_{\mathrm{ref}}$} --- the set against which novelty was judged: the entire Materials Project~\cite{Jain2013} together with the ${\sim}5$~million structures in the Alexandria database~\cite{schmidt2023alexandria}. A candidate was counted as novel only if it matched nothing in $\mathcal{D}_{\mathrm{ref}}$, which included the seed; thus, rediscovering a seed structure earned no credit.
\end{itemize}

\paragraph{Experiment 1: translation fidelity}
 
Experiment~1 tests the translation between crystal and text in both directions: from a crystal to its motif profile, and from that profile back to a crystal. Each of the $948$ seed structures was translated in both directions (Methods~2): Structure Translate wrote its motif profile, Motif Translate rebuilt a structure from that profile alone, and the profile was revised, up to five times, until the rebuilt structure passed the consistency tests of Methods~2. Each final reconstruction, as generated and without relaxation by DFT, was then compared with its original, the DFT-relaxed Materials Project structure, in three respects: the match rate, by pymatgen's \texttt{StructureMatcher} at the standard tolerance under which generative models report match rates (\texttt{ltol}~0.3, \texttt{stol}~0.5, \texttt{angle\_tol}~$10^\circ$); geometric and shape fidelity, by the root-mean-square displacement (RMSD) and the continuous shape measure (CShM)~\cite{pinsky1998cshm,zabrodsky1992csm}; and the local atomic environments, by the orbital-field matrix (OFM)~\cite{Pham2017,Pham2020}; the measures are defined in Supplementary Information~S4.2.

\paragraph{Experiment 2: design benchmark}
 
Experiment~2 ran the full loop for four generations. Each generation produced 200 proposals, of which the ERS, ranking on stability (Methods~5), selected 100 for DFT validation; the run length and both budgets were fixed in advance by the available DFT resources. Therefore, four generations resulted in a proposal pool of 800 and benchmark set of 400 validated candidates. The remaining 400 proposals, which the ERS had not selected, were relaxed after the run had completed and excluded from the hull reference of the benchmark; comparing the selected and unselected halves of the same pool indicates how much of the benchmark performance comes from the recommender and how much from motif editing.
 
MatEvolve was compared with two recent generative models: MatterGen~\cite{mattergen2025}, in both its magnetization-conditioned and unconditioned (Materials Project~+~Alexandria) forms, and the unconditioned Chemeleon2~\cite{chemeleon2025}. Every method received the same budget of 400 DFT-validated candidates, drawn from a disclosed proposal pool by its own ranking procedure: each baseline generated 1{,}000 candidates by its native sampling procedure, ranked them with CHGNet based on formation energy and predicted magnetic moment, and had its top 400 candidates validated.

\paragraph{Evaluation metrics}
 
Experiment~2 scored every method, and both MatEvolve halves, on the same four rates, each reported as a number of candidates and as a percentage of the method's 400 relaxations; where two methods come close, a footnote gives Fisher's exact test for their difference. The \SUN\ rate, the standard criterion of the structure-generation literature~\cite{xie2022cdvae,mattergen2025}, counts candidates that are \emph{stable} (energy above the convex hull $E_\mathrm{hull}\le0.1$~eV/atom, that is, on the hull or metastable within $0.1$~eV/atom of it), \emph{unique} (within the method's own set), and \emph{novel} (against $\mathcal{D}_{\mathrm{ref}}$). The design target, a rare-earth-lean magnet compound beyond the known structure types, adds two requirements, assessed separately. The \emph{magnetic rate} counts the \SUN\ candidates with a saturation magnetization $\mu_0 M_s$ above $1$~T, a value of permanent-magnet relevance. The \emph{new-prototype rate} counts the \SUN\ candidates whose AFLOW prototype label~\cite{Mehl2017,Hicks2021}---anonymized composition, Pearson symbol, space group and Wyckoff positions---does not appear in $\mathcal{D}_{\mathrm{ref}}$; therefore, substituting elements on a known lattice is invalid. Because the design target asks for both at once, a fourth rate---the \emph{joint rate}---counts the \SUN\ candidates that are new prototypes and magnetic at the same time.

\subsection{From crystal to motif profile and back}

\begin{figure*}[htbp]
\centering
\includegraphics[width=0.95\textwidth]{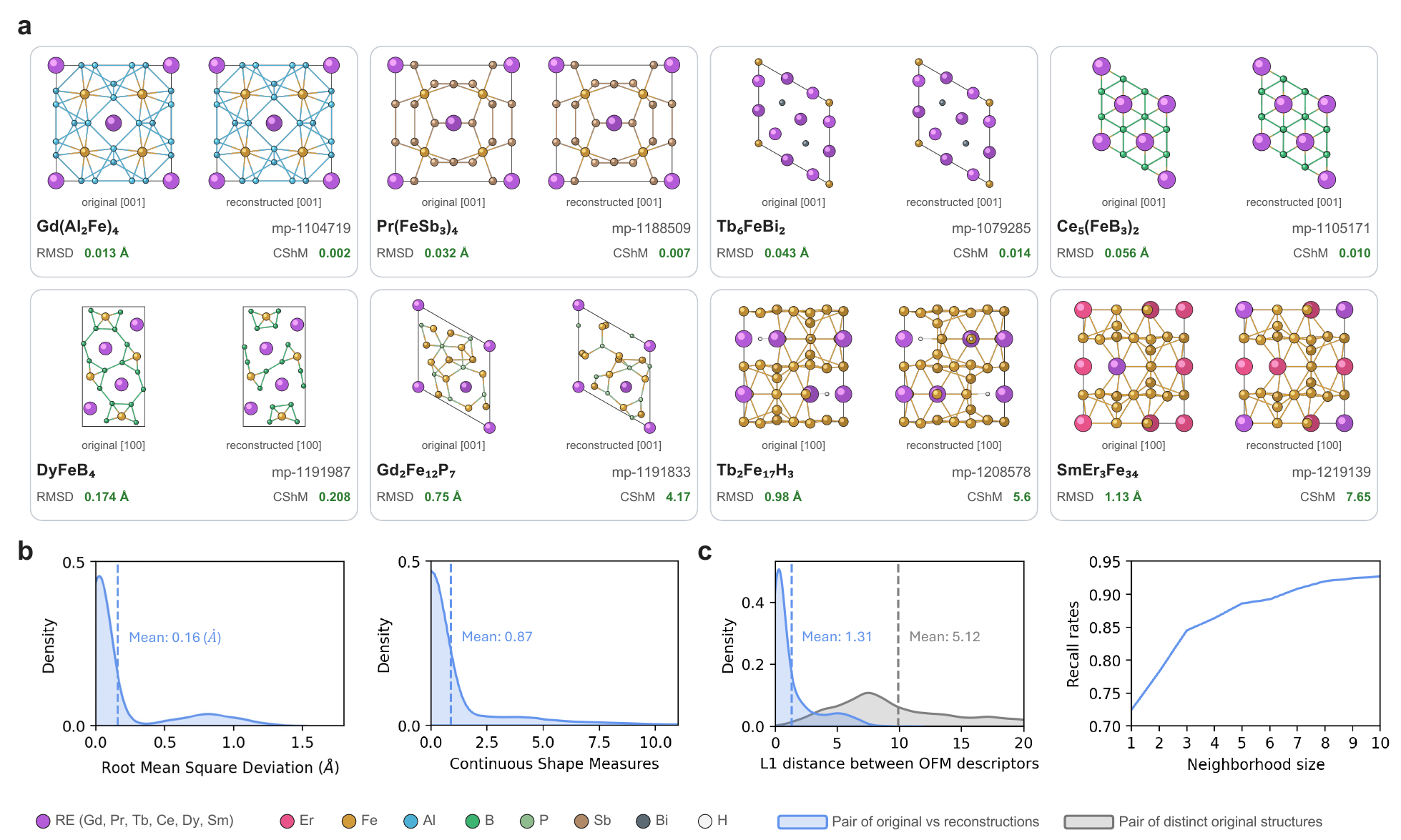}
\caption{\textbf{Fidelity of the translation from crystal to motif profile and back.} (a) Eight representative reconstructions spanning the range of outcomes, each viewed along [001] or [100] as labeled. In each pair, the left denotes the original Materials Project structure, and the right denotes its reconstruction by Motif Translate from the motif profile. RMSD (\AA) is the root-mean-square displacement between corresponding atoms after optimal superposition, and CShM the dimensionless continuous shape measure ($0$--$100$) of the reconstruction relative to the original~\cite{pinsky1998cshm,zabrodsky1992csm}. (b) Distributions of the RMSD (left) and CShM (right) over the matched reconstructions; dashed lines mark the means. (c) Fidelity of the local atomic environments, measured with the OFM~\cite{Pham2017,Pham2020}. Left: distribution of the $L_1$ distance between the OFM descriptors of each original and its reconstruction (blue; all original--reconstruction pairs of $\mathcal{D}_{\mathrm{seed}}$), against the distribution of $L_1$ distances from each original to its $100$ nearest neighbors among the other originals of $\mathcal{D}_{\mathrm{seed}}$ (gray); dashed lines mark the means. Right: fraction of originals whose own reconstruction is among their $k$ nearest OFM neighbors (recall), as a function of $k$; neighbors are sought among all other structures, that is, the remaining originals and all reconstructions.}
\label{fig:reconstruction}
\end{figure*}
 
We first tested the translation on which the framework rests in both directions: every crystal of $\mathcal{D}_{\mathrm{seed}}$ was written as a motif profile by Structure Translate, and a structure was rebuilt from that profile alone by Motif Translate. Because the profile withholds the coordinates, lattice parameters, space group, and structure-type label, the question is whether what remains suffices to recover the crystal. Of the $948$ reconstructions of $\mathcal{D}_{\mathrm{seed}}$ (Section~2.2), $59\%$ matched their originals at the tolerances conventionally used to evaluate deep-learning structure generators. Those generators report match rates of approximately $34\%$ to $64\%$ under the same settings on MP-20, a standard benchmark of Materials Project structures with at most 20 atoms per cell~\cite{xie2022cdvae,jiao2023diffcsp,miller2024flowmm}; the lower end is set by autoencoder reconstruction, the task closest to ours~\cite{xie2022cdvae}. The rate lies within that range; however, the comparison is only indicative: our reconstructions benefit from up to five profile revisions but involve cells of up to $100$ atoms, five times the MP-20 limit.

Figure~\ref{fig:reconstruction}a shows eight reconstructions beside their originals, spanning the range of outcomes. Four were rebuilt almost exactly: the ThMn$_{12}$-type Gd(Al$_2$Fe)$_4$, the filled skutterudite Pr(FeSb$_3$)$_4$, Tb$_6$FeBi$_2$, and the layered Ce$_5$(FeB$_3$)$_2$, with RMSDs of $0.013$--$0.056$~\AA. The boride DyFeB$_4$ was displaced further, by $0.174$~\AA, yet reproduced the boron network of its original. The three largest departures (Gd$_2$Fe$_{12}$P$_7$, the 2:17-type hydride Tb$_2$Fe$_{17}$H$_3$, and SmEr$_3$Fe$_{34}$) have RMSDs of $0.75$, $0.98$, and $1.13$~\AA, respectively: their atoms sit visibly off the original positions, but each reconstruction still contains the motifs of its original, as the motif-coverage test of Structure Translate requires.
 
These displacements measure mainly what the profile leaves free. A motif profile names the motifs a crystal contains, how they are arranged, and the bond-length regime of each, but gives no numerical bond length, coordinate, or lattice parameter. Therefore, Motif Translate must place the atoms, and its placement need not match the original. Scoring was also conservative: reconstructions were taken as Motif Translate emitted them, whereas the originals are DFT-relaxed; thus, part of each displacement may reflect relaxation alone. Over the matched reconstructions, the median RMSD was $0.025$~\AA\ and the mean $0.16$~\AA\ (Fig.~\ref{fig:reconstruction}b, left). Because the RMSD is an absolute measure, we also report a size-normalized measure of shape.
 
The continuous shape measure (CShM) scores the deviation of one structure from another after optimal superposition, on a scale from $0$, for an identical match, to $100$; it normalizes the squared displacement by the spread of the point set, so that structures of different size are compared on one scale (Supplementary Information~S4.2). Coordination chemistry reads it in bands: essentially ideal below approximately $0.1$, slightly distorted up to $3$, notably distorted from $3$ to $7.5$, and a breakdown of the target geometry is considered beyond $7.5$~\cite{pinsky1998cshm,alvarez2005shapemaps,shao2012alpo,funes2017butterfly}. Those bands were calibrated on single coordination polyhedra. Because a unit cell is more extended than a polyhedron, a given displacement yields a smaller CShM here; therefore, we use the bands to describe the distribution rather than to judge it. Across the matched reconstructions, $82\%$ fell in the essentially ideal band and a further $7\%$ in the slightly distorted band; $8\%$ showed notable distortion and $3\%$ lay beyond the breakdown boundary. The median CShM was $0.0054$ and the mean was $0.87$ (Fig.~\ref{fig:reconstruction}b, right), the mean was set by the tail. Over the reconstructions of $\mathcal{D}_{\mathrm{seed}}$ as a whole, not only the matched ones, the means rise to $0.43$~\AA\ for the RMSD and $3.94$ for the CShM. The eight examples of Fig.~\ref{fig:reconstruction}a place the bands on real structures: $0.002$--$0.014$ for the four rebuilt almost exactly, $0.21$ for DyFeB$_4$, and $4.17$, $5.60$, and $7.65$ for the three largest departures, the last lying just beyond the breakdown boundary.

The rebuilt structures also recover the local atomic environments of each crystal---the coordination of each atom, which governs stability, magnetic moment, and anisotropy. We compared them with the OFM~\cite{Pham2017,Pham2020}, which encodes this coordination, needs no atom correspondence, and applies to all $948$ original--reconstruction pairs. The median $L_1$ distance between the OFM descriptors of an original and its reconstruction was $0.44$ (mean $1.31$), against a median of $8.66$ (mean $9.95$) from each original to its $100$ nearest neighbors among the other originals of $\mathcal{D}_{\mathrm{seed}}$ (Fig.~\ref{fig:reconstruction}c, left). The reconstructions are also specifically close: for $72.5\%$ of the originals, the single nearest neighbor among all other structures---the remaining originals and every reconstruction---was their own reconstruction, and for $92.7\%$ it lay among the 10 nearest (Fig.~\ref{fig:reconstruction}c, right). Because the OFM also encodes chemical identity, part of this specificity reflects the shared composition rather than structure alone.
 
This experiment is the proof of concept on which the rest of the framework rests, in both directions: a crystal can be written as a motif profile, and that profile alone suffices to build the crystal back. Carrying no coordinates, lattice parameters, space group, or structure-type label, the profile still captures what the crystal is built from---its motifs, their arrangement and their coordination. Thus, the profile is not merely a description of a crystal but a specification from which one can be built: editing it specifies an offspring's motifs rather than its coordinates, and the placement is settled downstream by M3GNet pre-relaxation (Supplementary Information~S2) and first-principles relaxation.
 
\begin{table*}[t]
\centering
\caption{\textbf{Benchmark of MatEvolve against generative models on magnetic-material design.} Each method received 400 converged DFT relaxations, spent on the candidates its own ranking selected: for each baseline, the top 400 of 1{,}000 samples by CHGNet formation energy and magnetic moment; for MatEvolve, the top 100 of 200 proposals in each of four generations, ranked by the ERS. The second row holds the 400 MatEvolve proposals the ERS did not select, relaxed with the same DFT protocol. \SUN: stable ($E_\mathrm{hull}\le0.1$~eV/atom), unique within the set, and novel with respect to the Materials Project and Alexandria. Magnetic: \SUN\ candidates with $\mu_0 M_s>1$~T. New prototype: \SUN\ candidates whose prototype is absent from both databases. Both: \SUN\ candidates that are at the same time new prototypes and magnetic, the two requirements of the design target taken together. Bold text indicates the highest rate in each column.}
\label{tab:benchmark}
\small
\begin{tabular}{lcccc}
\hline
\textbf{Method} & \textbf{\SUN} & \multicolumn{3}{c}{\textbf{\SUN\ candidates meeting the design target}} \\
\cline{3-5}
 & \textbf{candidates} & \textbf{Magnetic} & \textbf{New prototype} & \textbf{Both} \\
\hline
\quad MatEvolve (selected by the ERS) & \textbf{115 (28.8\%)} & \textbf{104 (26.0\%)} & \textbf{89 (22.3\%)} & \textbf{80 (20.0\%)} \\
\quad MatEvolve (not selected by the ERS) & 97 (24.3\%) & 80 (20.0\%) & 53 (13.3\%) & 48 (12.0\%) \\
\hline
\quad Chemeleon2 (unconditioned) & 82 (20.5\%) & 5 (1.3\%) & 11 (2.8\%) & 3 (0.8\%) \\
\quad MatterGen (unconditioned) & 112 (28.0\%) & 8 (2.0\%) & 22 (5.5\%) & 2 (0.5\%) \\
\quad MatterGen (magnetization-conditioned) & 94 (23.5\%) & 88 (22.0\%) & 5 (1.3\%) & 5 (1.3\%) \\
\hline
\end{tabular}
\vspace{2mm}\\
\footnotesize
\textit{Note:} Each cell gives the number of candidates that meet the criterion, followed in parentheses by that number as a percentage of the method's 400 relaxations, that is, the rate. Novelty is decided by structure matching against all Materials Project and Alexandria entries (pymatgen \texttt{StructureMatcher}, database de-duplication tolerances); prototype status by the AFLOW label~\cite{Mehl2017,Hicks2021} of the symmetrized structure, matched against the labels of all such entries at a common symmetry tolerance ($0.1$~\AA).
\end{table*}

\subsection{Benchmarking and discovery of new stable magnets}

MatEvolve was benchmarked against two generative models, MatterGen~\cite{mattergen2025} and Chemeleon2~\cite{chemeleon2025}, under the protocol of Section~2.2. The ranking score of each method---the ERS for MatEvolve, CHGNet formation energy, and magnetic moment for the baselines---decided only which candidates were validated by DFT. Every rate below was determined from the DFT-relaxed structures and is given as a percentage of the 400 relaxations per method (Table~\ref{tab:benchmark}). Four rates were compared. The \SUN\ rate counts the candidates that are stable ($E_\mathrm{hull}\le0.1$~eV/atom), unique, and novel with respect to the Materials Project and Alexandria. The magnetic rate counts the \SUN\ candidates with $\mu_0 M_s>1$~T, and the new-prototype rate those whose prototype is absent from both databases; these two express the two requirements of the design target, and the joint rate counts the candidates that meet both at once.

Table~\ref{tab:benchmark} shows the four rates of each method: on the \SUN\ and magnetic rates MatEvolve is matched by one of the baselines, whereas on the new-prototype and joint rates it leads all of them clearly. On the \SUN\ rate, MatEvolve's 115 candidates ($29\%$) and unconditioned MatterGen's 112 ($28\%$) are statistically indistinguishable,\footnote{MatEvolve produced 3 more \SUN\ candidates than unconditioned MatterGen (115 against 112, out of 400 each). At this sample size, a difference of 3 is not statistically significant (two-sided Fisher's exact test, $p=0.88$), so neither method is shown to be higher.} with magnetization-conditioned MatterGen at $24\%$ and Chemeleon2 at $21\%$. Among these \SUN\ candidates, MatEvolve has 104 magnet compounds ($26\%$ of its 400 relaxations) and conditioned MatterGen has 88 ($22\%$), a gap not resolved at this sample size,\footnote{MatEvolve produced 16 more magnetic candidates than conditioned MatterGen (104 against 88, out of 400 each). At this sample size, a difference of 16 is not statistically significant (two-sided Fisher's exact test, $p=0.21$), so MatEvolve is not shown to be higher.} so the two are comparable on magnet compounds; unconditioned MatterGen and Chemeleon2, trained for stability alone, have only 8 ($2\%$) and 5 ($1\%$). By contrast, 89 of MatEvolve's \SUN\ candidates ($22\%$) are new prototypes, against 22 ($6\%$), 11 ($3\%$), and 5 ($1\%$) for unconditioned MatterGen, Chemeleon2, and conditioned MatterGen, respectively; the baselines' \SUN\ candidates are mostly new compounds on known prototypes, as expected of generators trained to reproduce known crystals. Requiring both properties at once, as the design target does, separates the methods most sharply: 80 of MatEvolve's candidates ($20\%$) are both magnetic and new prototypes, against 5 ($1.3\%$) for conditioned MatterGen, 3 ($0.8\%$) for Chemeleon2, and 2 ($0.5\%$) for unconditioned MatterGen---sixteen times the best baseline.
 
\begin{figure*}[t]
\centering
\includegraphics[width=0.95\textwidth]{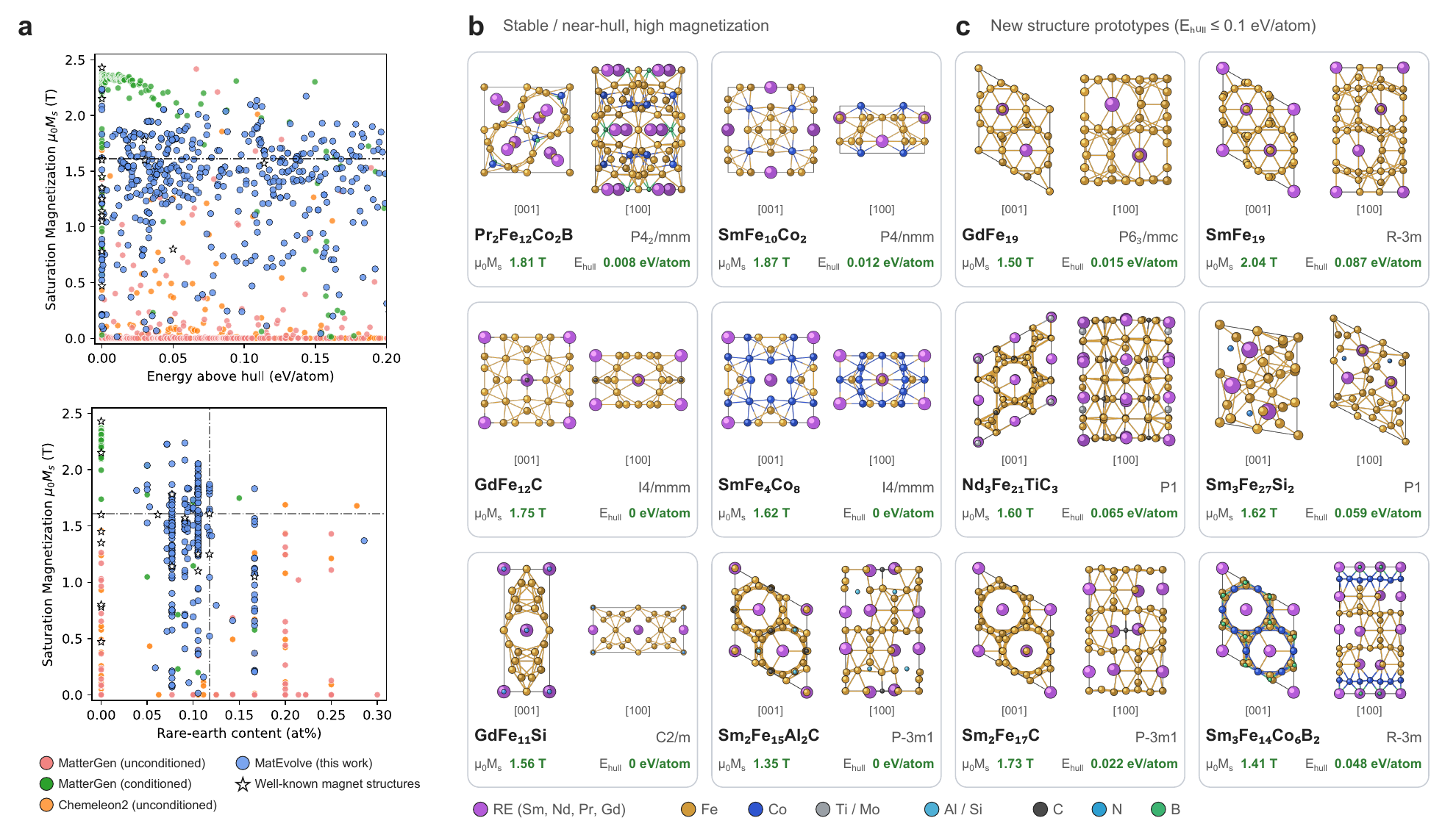}
\caption{\textbf{Magnetization, stability, and rare-earth content of the validated candidates, with representative MatEvolve structures.} (a) Saturation magnetization $\mu_0 M_s$ versus energy above the convex hull for the DFT-validated candidates of each method (top) and versus the rare-earth atomic fraction for the (meta)stable ones only, with $E_\mathrm{hull}\le0.1$~eV/atom (bottom): MatEvolve (blue), unconditioned MatterGen (pink), magnetization-conditioned MatterGen (green), and Chemeleon2 (orange). Gray stars: well-known magnet compounds, taken from the Materials Project. Dash-dot lines: $\mu_0 M_s$ and rare-earth fraction of Nd$_2$Fe$_{14}$B. (b) Six structures of known prototypes generated by MatEvolve, each with $\mu_0 M_s>1$~T and lying on or near the hull. (c) Six structures generated by MatEvolve that are both new prototypes and magnetic ($\mu_0 M_s>1$~T; $0<E_\mathrm{hull}\le0.1$~eV/atom). In (b) and (c), each DFT-relaxed structure is viewed along [001] and [100] and labeled with its formula, space group, $\mu_0 M_s$, and $E_\mathrm{hull}$.}
\label{fig:pareto}
\end{figure*}

 MatEvolve's rates combine two contributions: the motif editing that proposes candidates and ERS that selects which of them to validate. To separate the two, the 400 proposals the ERS did not select were relaxed after the run with the same DFT protocol (Table~\ref{tab:benchmark}, second row). Of these, 97 ($24\%$) are \SUN, 53 ($13\%$) are new prototypes, and 80 ($20\%$) are magnetic, against 115, 89, and 104 for the selected set, respectively; the largest difference is in new prototypes ($13\%$ against $22\%$). The ERS ranks on stability alone; thus, the gain in stable candidates (115 against 97) is its expected effect. Because the other two rates count subsets of the stable candidates, their gain may follow from it rather than from any preference of the ERS for new prototypes or magnet compounds. What the nonselected proposals show is that motif editing alone, before any selection, already yields new prototypes at $13\%$, more than twice the best baseline, and magnet compounds at $20\%$, level with conditioned MatterGen; the ERS then raises both yields further.

Beyond these rates, Figure~\ref{fig:pareto}a examines the validated candidates further, by magnetization, stability, and rare-earth content. Only MatEvolve and conditioned MatterGen produce high-magnetization candidates in appreciable numbers (upper panel): the unconditioned baselines lie mostly below $0.5$~T, and the few Chemeleon2 candidates near or above $1$~T are rare-earth-rich (fractions of $0.17$--$0.30$). Conditioned MatterGen reaches the highest values, up to about $2.4$~T near the hull; MatEvolve follows, with many of its candidates above the $1.6$~T of Nd$_2$Fe$_{14}$B and the highest at $2.2$~T. The lower panel keeps only the (meta)stable candidates ($E_\mathrm{hull}\le0.1$~eV/atom) and plots them against the rare-earth fraction, which matters because a permanent magnet needs some rare earth---its $4f$ electrons supply the magnetocrystalline anisotropy that makes the magnet hard---but as little as possible because rare-earth materials are scarce and costly. Here, the two methods part: conditioned MatterGen's high magnetizations lie almost entirely at zero rare-earth content, in Fe--Co-rich compositions similar to the soft magnetic materials $\alpha$-Fe and Fe$_{65}$Co$_{35}$; however, all of MatEvolve's (meta)stable candidates above $1$~T contain rare earth, and $88\%$ of them lie at fractions up to $0.12$, the range of the well-known magnet compounds SmFe$_{11}$Ti and Sm(Fe,Co)$_{12}$ ($0.08$), Sm$_2$Fe$_{17}$N$_3$ ($0.09$), and Nd$_2$Fe$_{14}$B ($0.12$).

Figure~\ref{fig:pareto}b shows six candidates that exceed $1$~T, lie on or near the hull, and are new to $\mathcal{D}_{\mathrm{ref}}$ as ordered compounds but not as prototypes. All belong to the principal Fe-rich families studied for rare-earth permanent magnets~\cite{coey1990}, each represented in $\mathcal{D}_{\mathrm{seed}}$, so they test physical plausibility rather than demonstrate discovery. Four are ThMn$_{12}$-type 1:12 derivatives: GdFe$_{12}$C, with interstitial carbon ($2.22$~T); GdFe$_{11}$Si, with Si substitution ($2.05$~T); and SmFe$_{10}$Co$_2$ and SmFe$_4$Co$_8$ ($1.87$ and $1.62$~T), ordered variants of the known Sm(Fe,Co)$_{12}$ system with Fe and Co on distinct sites. The other two are Pr$_2$Fe$_{12}$Co$_2$B, a Co-substituted Nd$_2$Fe$_{14}$B-type 2:14:1 phase ($1.81$~T), and Sm$_2$Fe$_{15}$Al$_2$C, a Th$_2$Zn$_{17}$-related 2:17 carbide with Al substitution ($1.35$~T). The interstitial and substituent elements (C, Co, Si, Al) are those used in practice to stabilize or tune these families.

Figure~\ref{fig:pareto}c shows six candidates that are metastable ($0<E_\mathrm{hull}\le0.1$~eV/atom), exceed $1$~T, and realize a prototype absent from $\mathcal{D}_{\mathrm{ref}}$. They reach their new prototypes in two ways. The first is \emph{dumbbell substitution}, the mechanism behind the known Fe-rich families: replacing a fraction $s$ of the R atoms of CaCu$_5$-type RT$_5$ by Fe--Fe dumbbells gives the R$_{1-s}$Fe$_{5+2s}$ series, of which R$_2$Fe$_{17}$ ($s=1/3$) and ThMn$_{12}$-type RFe$_{12}$ ($s=1/2$) are established members~\cite{coey1990}. Three candidates extend this series to members absent from $\mathcal{D}_{\mathrm{ref}}$. GdFe$_{19}$ ($1.82$~T) and SmFe$_{19}$ ($2.04$~T) lie at $s=2/3$, beyond RFe$_{12}$; they adopt the space groups of the two R$_2$Fe$_{17}$ polytypes ($P6_3/mmc$ and $R\bar{3}m$) and contain isolated Fe--Fe pairs at $2.36$--$2.39$~\AA, the dumbbell distance of the 2:17 phases. Sm$_3$Fe$_{27}$Si$_2$ ($1.62$~T), triclinic, has the 3:29 stoichiometry ($s=2/5$) with Si on transition-metal sites and likewise contains isolated pairs. The second approach is \emph{motif substitution}: a motif of a known structure is replaced by, or supplemented with, a different one. Sm$_2$Fe$_{17}$C ($1.73$~T) fills every third 2:17 block with carbon, as Sm$_2$Fe$_{17}$N$_3$ is filled with nitrogen~\cite{Miyake18}, the carbon occupying the same octahedral sites of four Fe and two Sm. Nd$_3$Fe$_{21}$TiC$_3$ ($1.60$~T) places carbon in similar octahedral interstices within a Nd--Fe--Ti framework. Sm$_3$Fe$_{14}$Co$_6$B$_2$ ($1.41$~T) inserts boron-centered Co$_6$ trigonal prisms, the Co analog of the Fe$_6$B prisms of Nd$_2$Fe$_{14}$B~\cite{Herbst1991}, into an Fe--Co framework. In all six, the local environments---dumbbells, C- and B-centered sites---are known motifs or close variants of them; the novelty lies in their arrangement.

Sm$_2$Fe$_{17}$C also illustrates the degree to which two arrangements of one composition can differ. Sm$_2$Fe$_{17}$C$_x$ carbides are experimentally established~\cite{deMooij1988}, and the composition is already in $\mathcal{D}_{\mathrm{ref}}$ as mp-1219267, a monoclinic $C2/m$ structure with a different carbon arrangement, $0.28$~eV/atom above the hull at $0.064$~T; the generated $P\bar{3}m1$ structure lies $0.02$~eV/atom above the hull at $1.73$~T. Whether the difference stems from the carbon arrangement or from the magnetic states to which the two calculations converged cannot be ascertained from this pair alone.
 
Taken together, these results show that editing crystals in the language of motifs reaches both requirements of the target simultaneously, a new prototype that is also a magnet compound, at a rate that generators sampling the distribution of known crystals do not approach. It did so with the target written into the agents' prompts rather than trained into a model, which suggests that the same framework could be redirected to other design targets in materials discovery by restating the objective; however, this remains to be tested. In the new prototypes examined above, the novelty lies not in new motifs but in a new arrangement of known ones; the question we next address is whether that holds across all new-prototype candidates.

\subsection{Structural insights into novelty and stability through motif analysis}

\begin{figure*}[t]
\centering
\includegraphics[width=0.95\textwidth]{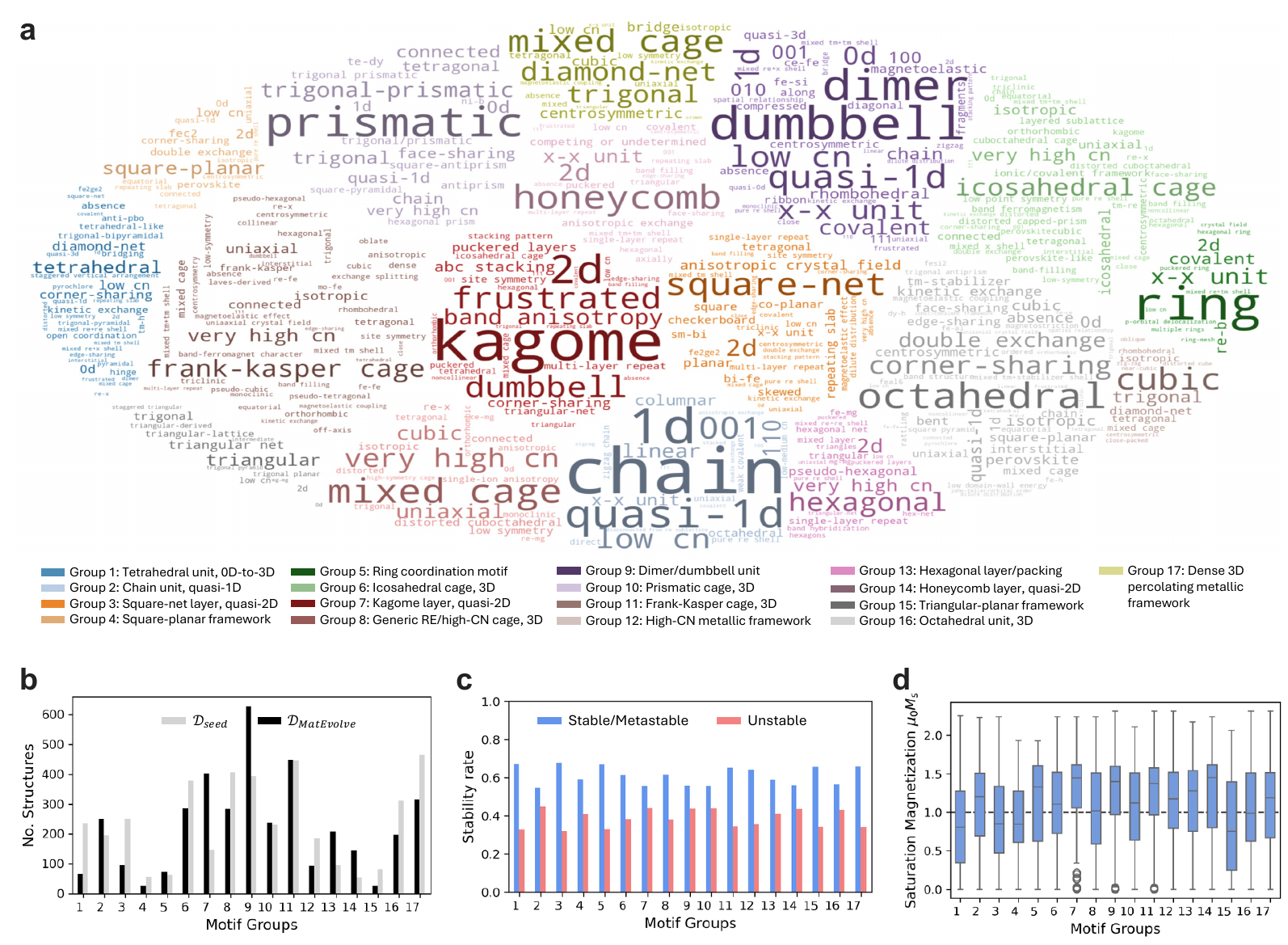}
\caption{\textbf{Motif groups discovered from the seed and assigned across the study.} The groups are discovered from the seed structures' motifs, and the motifs of the generated structures are then assigned to them (11{,}478 motifs across all 1{,}748 structures). (a) Word clouds of the 17 motif groups; each group is summarized by its most representative keywords, ranked by TF--IDF over the group's motif descriptions from candidate terms proposed by the language-model agents. (b) For each group, the number of contributing structures from the Materials Project seed set ($\mathcal{D}_{\mathrm{seed}}$, gray) and from MatEvolve ($\mathcal{D}_{\mathrm{MatEvolve}}$, black). (c) For each group, the fraction of contributing structures that are stable or metastable (blue) versus unstable (red). (d) For each group, the distribution of saturation magnetization $\mu_0 M_s$ across the structures that contain a motif from that group; the dashed line marks the $1$~T design target.}
\label{fig:motifmap}
\end{figure*}

The interpretable motif profile allows us to return to the question Sm$_2$Fe$_{17}$C posed and to characterize the discoveries, not merely count them: for the structures the run produced---all 800, novel or not---we can ask where the novelty of the new prototypes resides and what underlies their stability. Both questions turn on the local motifs from which the structures are built; therefore, we first establish a common vocabulary of geometric groups from the seed alone. A consensus procedure---five language-model voters casting categorical votes, with a deterministic routine performing all merging, splitting, and counting, so that the number of groups is discovered rather than fixed (Methods~4)---partitions the motifs of the seed structures into 17 groups. Each motif in the generated structures is then assigned to one of these groups using the same categorical voting procedure, or, if it matches none, it is left unassigned. This shared labeling allows the recommender to score each candidate against the stability statistics derived from the seed.

\begin{figure*}[t]
\centering
\includegraphics[width=\textwidth]{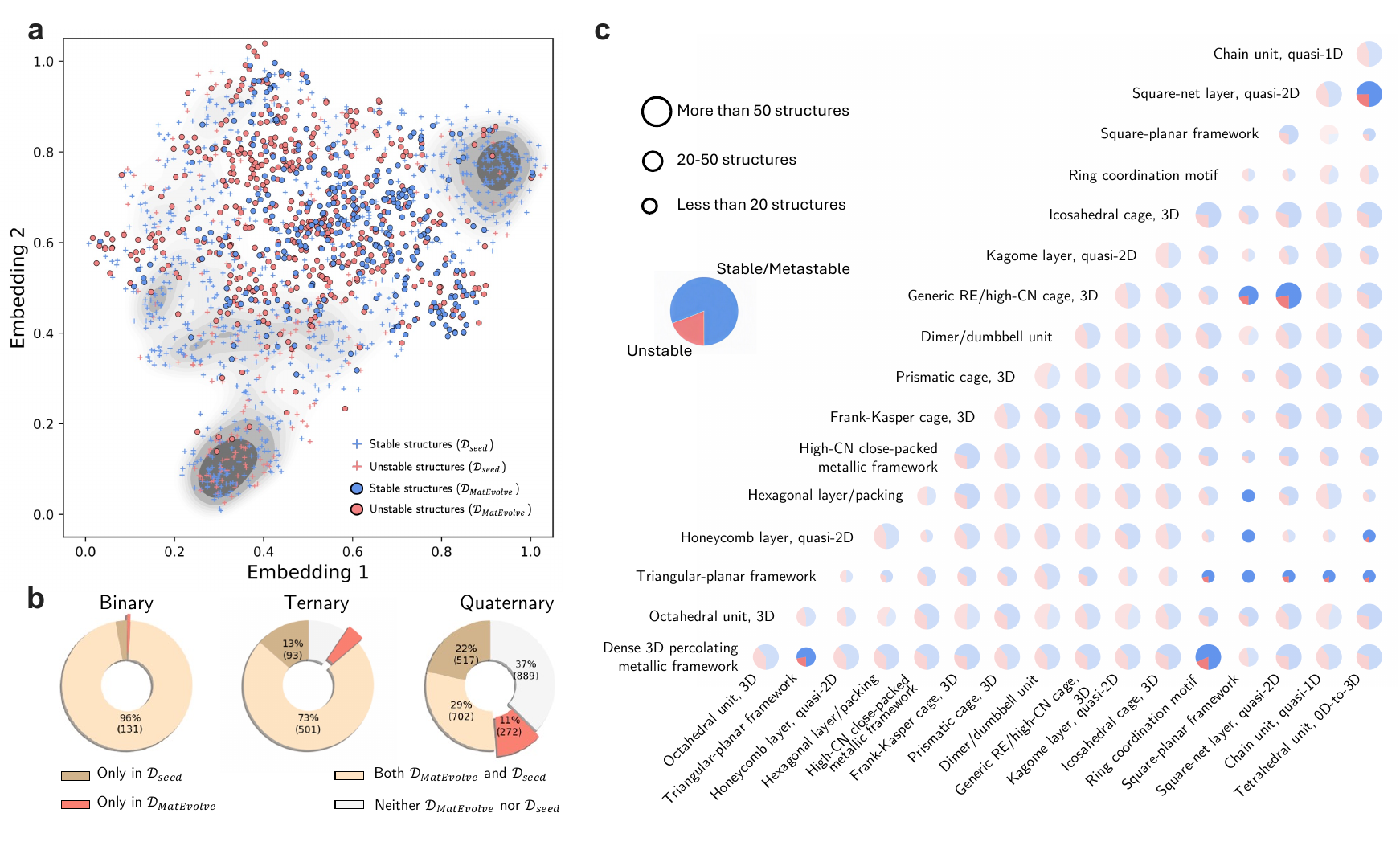}
\caption{\textbf{Motif-group composition, combination coverage and pairwise stability across the 1{,}748 structures.} (a) Structure map: each of the 1{,}748 structures is represented by the set of 17 motif groups it contains and embedded so that structures sharing groups---or built from mutually substitutable groups---lie close together (t-SNE on a substitutability-weighted set-overlap distance; axes are arbitrary embedding coordinates). Marker shape encodes origin---Materials Project seed structures (crosses) versus MatEvolve-generated structures (circles). The origin is denoted by marker shape, not by shade as in Fig.~\ref{fig:motifmap}b, and the color encodes stability: blue for stable or metastable and red for unstable; gray contours are a kernel-density estimate of the seed structures' distribution in the embedding. (b) Coverage of motif-group combinations by order: for the binary, ternary, and quaternary combinations of the 17 groups, the fraction of all possible combinations observed only in the seed set ($\mathcal{D}_{\mathrm{seed}}$), only among the generated structures ($\mathcal{D}_{\mathrm{MatEvolve}}$), in both, or in neither (counts in parentheses). (c) Pairwise motif-group stability: for every pair of the 17 groups, a pie denotes the fraction of the structures containing the pair that is stable or metastable (blue) versus unstable (red), and pie size is binned by the number of contributing structures (more than 50, 20--50, or fewer than 20); pairs with a stable-or-metastable fraction of at least $75\%$ are shown in full color and the others faded.}
\label{fig:combimap}
\end{figure*}

Across the study, the 11{,}478 motifs from all 1{,}748 structures were distributed across the 17 groups, apart from a small unassigned tail noted below, each of which was summarized as a word cloud in Fig.~\ref{fig:motifmap}a by its most representative motif keywords (construction detailed in Supplementary Information~S6). The groups were defined along the axes used by a crystal chemist---unit class, coordination geometry, and coordination-number band---rather than in a latent coordinate. Most groups represented a specific geometry, spanning much of the standard vocabulary of intermetallic chemistry---tetrahedral, octahedral, and dimer units; triangular-planar and square-planar frameworks; chain, square-net, kagome, honeycomb, hexagonal, and ring motifs; and icosahedral, Frank--Kasper, prismatic, and generic cages. The emergence of such a small, recognizable set was not an artefact of clustering: intermetallic coordination is known to collapse onto approximately 20 recurring environments---with the icosahedron and $Z12$--$Z16$ Frank--Kasper cages among the most common~\cite{daams1992aet,villars1991atlas,frank1958,frank1959}---which our clustering recovered. 

Every group was broadly compatible with stability: across all 17 groups, the stable-or-metastable fraction fell within a similar range ($55$--$68\%$; Fig.~\ref{fig:motifmap}c); thus, no single geometry was uniquely stabilizing or destabilizing. Whether a structure was stable thus appeared to depend more on how its groups \emph{combined} than on which groups it contained, which we analyzed further below. By contrast, saturation magnetization separated the groups (Fig.~\ref{fig:motifmap}d): the highest values belonged to the kagome layer, dimer/dumbbell unit, Frank--Kasper cage, and honeycomb layer, while the dense metallic frameworks and hexagonal layer also averaged above the $1$~T target. These motifs marked iron-rich environments, and the ordering tracked iron content: the transition-metal nets and dense metallic frameworks carried the itinerant $3d$ moment~\cite{yin2022kagome,coey1990}, while the dimer/dumbbell and Frank--Kasper-cage groups showed high values because they marked iron-rich rare-earth--iron intermetallics---the cage housing the rare-earth atom that supplies the $4f$ anisotropy~\cite{Miyake18}, with the moment coming from the surrounding iron framework. Recovering this magnetically meaningful vocabulary from geometry alone made the partition interpretable.

The vocabulary of the seeds accounted for the great majority of the motifs in the generated structures. Assigning each generated motif to one of the 17 seed groups---using the same categorical vote, which could also return \emph{none}---mapped most of them onto an existing group, so that every group was used by both the seed and generated set (Fig.~\ref{fig:motifmap}b). A small fraction was instead returned unassigned---local geometries with no counterpart among the seed's groups---and were set aside from the combinatorial analysis below. The small size of this tail was important: MatEvolve's structural novelty was predominantly combinatorial, arising far more from new arrangements of the shared vocabulary than from new building blocks, as the next paragraphs quantified.

We extended this analysis to the whole-structure level by representing each structure as the set of motif groups to which its motifs were assigned. Fig.~\ref{fig:combimap}a maps the 1{,}748 structures under this motif-group representation---structures sharing groups, or mutually substitutable ones~\cite{Ha2021,Pham2020}, were placed close together (details of the map construction are given in Supplementary Information~S5). The seed structures formed two dense regions (gray kernel-density contours), and the generated structures largely gathered along the border between them rather than in a region isolated from the seed. A structure combining groups from both regions was placed between them; therefore, this bridging position reflected new combinations of the existing groups. Since almost every generated motif still mapped onto a seed group rather than remaining unassigned, the generated structures recombined existing groups rather than introducing new ones.

Fig.~\ref{fig:combimap}b quantified this pattern by classifying the binary, ternary, and quaternary combinations of the 17 groups---the sets of two, three, and four groups---based on whether they appeared in the seed, the generated set, both, or neither. Of the binary and ternary combinations that occurred at all, almost all already appeared in the seed---only 1 of the 136 observed binary combinations and 30 of the 624 observed ternary combinations ($5\%$) were unique to the generated set. Therefore, at low orders, the generated set introduced few new combinations. The novelty appeared at the whole-structure level, as the quaternary analysis below shows. Quaternary combinations were sampled far more sparsely: of the 2{,}380 possible combinations, 702 occurred in both sets, 517 in the seed only, 272 in the generated set only, and 889 in neither. Of the 800 generated structures, 182 were built from combinations of motif groups that appeared in no seed structure, suggesting that---at the resolution of this grouping---MatEvolve's novelty was largely combinatorial, favoring new combinations of the seed's existing groups over new groups.

The stability of those combinations is a separate issue. For each of the 136 group pairs, we computed the fraction of structures containing that pair that is stable or metastable across the seed and generated sets (Fig.~\ref{fig:combimap}c). The highest fractions, up to $100\%$, belong to pairs found in fewer than 20 structures and are therefore not robust. Among the pairs supported by more than 50 structures, three stand out, at $76$--$82\%$. Each corresponds to the local-motif signature of a recognizable intermetallic family---an analogy we draw after ranking rather than a prediction: a ring coordination motif within a dense percolating metallic framework ($82\%$), as in CaCu$_5$-type RT$_5$ compounds and their dumbbell-substituted derivatives, the RFe$_{12}$ and R$_2$Fe$_{17}$ phases, in which the rare-earth atom sits in a hexagonal ring of transition-metal atoms~\cite{coey1990}; and a square-net layer paired with either a generic rare-earth high-coordination cage ($78\%$) or tetrahedral units ($76\%$), as in ThCr$_2$Si$_2$-type RT$_2$X$_2$ compounds (R = rare earth, T = transition metal, X = Si, Ge, P, or As), whose layers of edge-sharing TX$_4$ tetrahedra place the transition-metal atoms on a square net between rare-earth atoms in high-coordination cages.

Taken together, these analyses support two conclusions at this group resolution. First, MatEvolve advances mainly by recombination rather than invention: where its structures depart from the seed, they form new arrangements of the existing motif vocabulary rather than new building blocks, and the most stable of these fall into recognizable intermetallic families. This is not in contrast with the new prototypes reported above: a prototype is a whole-structure arrangement; thus, a combination of familiar local motifs realized in an arrangement absent from the reference can be combinatorially familiar in its parts while being new as a prototype---the advance is new global arrangements of a shared local vocabulary, not new local building blocks. Second, that vocabulary is human-readable; therefore, the representation that generates a structure also allows us to read it back---a step toward an interpretable design language. Turning these descriptive regularities into predictive design rules and probing novelty below the group resolution may be explored in future work.


\section*{Discussion}
\label{sec:discussion}

MatEvolve shows that the motif profile serves not merely as a description of a material but as the medium for material design: the agent edits the profile and constructs a crystal from the modified one, and the most promising candidates are validated by first-principles calculation. On the rare-earth-lean permanent-magnet task it identified novel (meta)stable crystals that are both new prototypes and magnet compounds an order of magnitude more often than generative models, given the same validation budget. It also exceeded every baseline several-fold on new prototypes and matched a magnetization-conditioned generator on magnet compounds without any training on the property. In this benchmark, an agent designing at the level of motifs is thus competitive with generators trained for the task, and exceeds them where the task calls for new structure types.

A second strength is its potential transferability. The agent is a commercial language model (Claude Fable~5) that was used as delivered and never fine-tuned on magnetic-material data; however, it designs competitive magnet compounds. The magnetism-specific ingredients are a few swappable components---the objective, property surrogate, seed set, and first-principles settings---while the motif-editing loop is not tailored to magnetism. Therefore, retargeting should require no retraining, only a restated objective and revised task procedures, carried out largely in natural language so that a scientist from the target field can direct the search. However, this remains to be demonstrated on a second target.

Considered more broadly, MatEvolve is a small instance of a larger shift---AI becoming an engine of scientific discovery, now argued to mark a new, fifth paradigm of science~\cite{berens2023,bishop2022ai4science,ioannidis2024fifthparadigm}. What a materials example adds is a template built on discipline rather than autonomy: an agent earns trust when its proposals are readable, admitted only once they reconstruct a real structure, and adjudicated by first principles rather than by the model itself. Its deeper promise is an iterative cycle between the two directions of this paradigm. One---\emph{AI for science}---is realized here: the agent accelerates discovery, proposing and verifying candidates at a scale beyond manual reach. The other---\emph{science for AI}---is only motivated by this work: because the agent reasons in readable terms, the insights it exposes may be read back to sharpen the scientist's understanding and inform the design of a subsequent, more capable agent---a loop we do not close here. This mutual reinforcement, rather than autonomy, is seen as the way forward: not a black box that supplants the scientist, but AI and science advancing each other.

Several limitations bound these claims. The demonstration is computational, and its property scope is narrow. DFT evaluates stability and saturation magnetization, but not magnetocrystalline anisotropy, the Curie temperature, or synthesizability---each decisive for a working magnet. Thus, the candidates are precursors to functional magnets rather than instances of them. The magnetization values are moreover $0$~K collinear upper bounds (for the heavy rare earths, the antiparallel $4f$--$3d$ coupling can partly compensate the transition-metal moment), and stability is assessed thermodynamically, from pure generalized gradient approximation (GGA) energies and without a phonon screen. Establishing whether any candidate becomes a usable magnet will require these further figures of merit and, ultimately, synthesis.

The search is also loosely guided, and its analysis inherits assumptions from the seed set. The agent itself sees no first-principles feedback within a design step---DFT labels re-enter only through the parent pool and ERS; the motif vocabulary is clustered once from the seed structures, so novelty is read categorically, as previously unobserved \emph{combinations} of fixed groups; and the recommender is calibrated on crystals only by analogy with the alloy setting in which it was validated. The natural next steps are feeding the surrogate and DFT signals back into the agent's reasoning, re-deriving the groups as data accrue, grading novelty continuously, and validating the recommender directly on crystals. A comparison of the motif profile with earlier text-based treatments of crystals is provided in Supplementary Information~S7.


\section*{Methods}
\label{sec:methods}

\subsection*{1. Evolutionary search over motif profiles}

MatEvolve combines the components above into an evolutionary search over motif profiles (Algorithm~\ref{alg:loop}). The population is initialized from the seed set $\mathcal{D}_{\mathrm{seed}}$ and evolved for four generations. In each generation, parents are selected from the current pool; Structure Translate writes each selected parent as a motif profile (Methods~2); Motif editing edits these parent profiles into offspring profiles, which Motif Translate realizes as candidate structures (Methods~3); the ERS ranks the candidates (Methods~5); and DFT is used to validate them in that priority order (Methods~6). Every validated candidate---both stable and unstable---is added to the pool, updating both the parent pool and ERS evidence set for the next generation. Thus, the search is informed by all of its own first-principles labels.

\emph{Parent selection.} In the magnet run reported herein, each design step draws two parents from the current pool---one from the $200$ structures with the lowest energy above the convex hull and one from the $200$ structures with the highest saturation magnetization---and combines them by crossover into a single offspring. Thus, each candidate inherits from a stability-favored and magnetization-favored parent. The element- and geometry-mutation operators (Methods~3; Supplementary Information~S3) are available in the framework but were not applied in this run. In each generation, $200$ offspring are proposed and the ERS selects the top $100$ for DFT validation, giving $400$ validated structures across the four generations; the $400$ unselected proposals are additionally relaxed as diagnostics, outside the search loop.

\begin{algorithm}[tbp]
\caption{MatEvolve generation loop}\label{alg:loop}
\footnotesize
\textbf{Input:} seed population $\Pi_0$; objective $\mathcal{P}$\\
\textbf{Output:} the DFT-validated candidates accumulated over all generations
\begin{pseudo}
\item \kw{for} $g=1$ \kw{to} $4$ \kw{do}
\item \ind select a parent set from $\Pi_{g-1}$
\item \ind $\mathcal{T}\gets\{\proc{StructureTranslate}(S):S\text{ a selected parent}\}$ 
\item \ind $\mathcal{T}_{\mathrm{off}}\gets\{\proc{MotifEditing}(\text{parents}\subseteq\mathcal{T},\,\mathcal{P})\}$
\item \ind $\mathcal{O}\gets\{\proc{MotifTranslate}(T_{\mathrm{off}}):T_{\mathrm{off}}\in\mathcal{T}_{\mathrm{off}}\}$
\item \ind rank $\mathcal{O}$ by the ERS \cmt{Details are shown in Methods~5}
\item \ind $\mathcal{V}\gets\proc{DFT\text{-}validate}(\text{top-}100\text{ of }\mathcal{O}\text{ by ERS rank})$
\item \ind $\Pi_g\gets\Pi_{g-1}\cup\mathcal{V}$
\item \kw{end for}
\end{pseudo}
\end{algorithm}

\subsection*{2. Structure Translate --- forward translation with bidirectional-consistency enforcement}

Structure Translate turns a crystal $S$ into a \emph{validated} motif profile $T$. A profile that merely sounds plausible can still be incomplete as it may omit details needed to specify the structure; therefore, a profile is retained only if the original crystal can be rebuilt from it. We call this round-trip test \emph{bidirectional consistency}. The translation is performed through the accept-or-revise loop of Algorithm~\ref{alg:translate} using the two prompt-based basis functions, \textsc{CIF2Text} and \textsc{Text2CIF}, whose implementations are given in Supplementary Information~S1 and Supplementary Information~S2. \textsc{CIF2Text} drafts a profile $T$ from $S$; \textsc{Text2CIF} reconstructs a crystal's $S'$ from $T$; and $S'$ is passed through \textsc{CIF2Text} once more to produce a profile $T'$ for checking. When the checks fail, the agent revises the profile rather than discarding it: deterministic, machine-checkable feedback---the target motifs that were missing or spurious, and any composition or symmetry mismatch---is returned, and a revised profile $T$ is drafted, with a maximum of five attempts.

\begin{algorithm}[tbp]
\caption{\proc{StructureTranslate}: forward translation with bidirectional-consistency enforcement}\label{alg:translate}
\footnotesize
\textbf{Input:} crystal $S$\\
\textbf{Output:} validated motif profile $T$, or \kw{reject}
\begin{pseudo}
\item $\textit{feedback}\gets\varnothing$
\item \kw{for} $i=1$ \kw{to} $5$ \kw{do}
\item \ind $T\gets\proc{CIF2Text}(S,\textit{feedback})$
\item \ind $S'\gets\proc{Text2CIF}(T)$
\item \ind $T'\gets\proc{CIF2Text}(S')$
\item \ind \kw{if} $\proc{Coverage}(T,T')$ \kw{and} $\proc{StructureMatcher}(S,S')$ \kw{then} \kw{return} motif profile $T$
\item \ind $\textit{feedback}\gets$ missing/spurious motifs; composition or symmetry mismatch
\item \kw{end for}
\item \kw{return} \kw{reject}
\end{pseudo}
\end{algorithm}

\emph{Acceptance --- round-trip fidelity.} A draft profile $T$ is accepted---and the loop stops---only when the crystal rebuilt from it agrees with the original in two complementary aspects: it must be built from the same motifs and must assemble them into the same structure. The two checks in Algorithm~\ref{alg:translate} test these requirements in turn. The first, \emph{motif coverage} \textsc{Coverage}$(T,T')$, asks whether the reconstruction retains the same building blocks; it compares the target motif set $T=\{t_1,\dots,t_{|T|}\}$ with the set $T'$ obtained by re-translating the reconstruction. By using $[\,t\preceq U\,]$ to denote that a motif $t$ is present in a set $U$ (matched at the archetype level, with a motif split into finer pieces counted as present), the recall $r$ and precision $p$ are as follows:
\begin{equation}
r=\frac{1}{|T|}\sum_{t\in T}\bigl[\,t\preceq T'\,\bigr],\qquad p=\frac{1}{|T'|}\sum_{t'\in T'}\bigl[\,t'\preceq T\,\bigr],
\end{equation}
Requiring $r=1$ ensures that every target motif reappears; thus, no building block is lost or misdescribed in the round trip; the looser $p\ge0.7$ allows a few additional motifs, since a faithful reconstruction may describe the same structure at slightly finer granularity. An excess of detail is therefore considered benign, whereas a missing motif represents lost information. The second check, \textsc{StructureMatcher}$(S,S')$, then asks whether those building blocks were assembled correctly, accepting the profile only when the rebuilt crystal $S'$ matches the original $S$ geometrically (pymatgen, using the acceptance tolerances \texttt{ltol}~0.4, \texttt{stol}~0.7, \texttt{angle\_tol}~$15^\circ$). The two checks are complementary as a profile can name the correct motifs but arrange them incorrectly, or reproduce the coordinates while relying on different motifs. Therefore, passing both certifies that $T$ encodes $S$ faithfully and sufficiently. Coverage is applied first because it directly tests the defining building blocks, providing a semantic agreement that a global coordinate metric can mask. The recall and precision defined here are reused wherever two motif profiles are compared (Methods~3). 

\subsection*{3. Motif editing and Motif Translate --- offspring proposal and realization}

Motif editing and Motif Translate together create a new candidate crystal: the first edits the motif profiles of existing structures---recombining or mutating their building blocks rather than sampling from a distribution of known structures---and the second realizes the edited profile as a crystal. Rebuilding a crystal from an edited profile is underdetermined because the same profile can be realized by more than one arrangement of atoms. Therefore, a generated crystal is retained only if it actually contains the motifs that were designed.

\emph{Motif editing.} A design agent, the evolutionary operator \textsc{EvoOperator} (Supplementary Information~S3), reads one or two parent profiles from the current pool and proposes an offspring profile $T_{\mathrm{off}}$ by crossing over motifs from two parents or by applying element or geometry mutations to a parent's motifs.

\emph{Motif Translate.} Motif Translate follows the loop in Algorithm~\ref{alg:design}: the basis agent \textsc{Text2CIF} (Supplementary Information~S2) realizes $T_{\mathrm{off}}$ as a candidate crystal $S_{\mathrm{off}}$, which is accepted or retried as follows.

\begin{algorithm}[tbp]
\caption{\proc{MotifTranslate}: realization of an offspring profile}\label{alg:design}
\footnotesize
\textbf{Input:} offspring profile $T_{\mathrm{off}}$\\
\textbf{Output:} candidate structure $S_{\mathrm{off}}$, or \kw{reject}
\begin{pseudo}
\item $\textit{feedback}\gets\varnothing$
\item \kw{for} $j=1$ \kw{to} $5$ \kw{do}
\item \ind $S_{\mathrm{off}}\gets\proc{Text2CIF}(T_{\mathrm{off}},\textit{feedback})$
\item \ind $T'_{\mathrm{off}}\gets\proc{CIF2Text}(S_{\mathrm{off}})$
\item \ind \kw{if} $\proc{Coverage}(T_{\mathrm{off}},T'_{\mathrm{off}})$ \kw{then} \kw{return} structure $S_{\mathrm{off}}$
\item \ind $\textit{feedback}\gets$ uncovered/spurious motifs
\item \kw{end for}
\item \kw{return} \kw{reject}
\end{pseudo}
\end{algorithm}

\emph{Acceptance --- realization consistency.} A generated crystal is accepted only if it realizes the profile from which it was designed. Here, the target is a \emph{profile}, not an existing crystal. Therefore, no reference structure exists to confirm a geometric match, and the check reduces to the motif-coverage criterion of Methods~2: \textsc{CIF2Text} re-derives the motifs of $S_{\mathrm{off}}$ as a profile $T'_{\mathrm{off}}$, and $S_{\mathrm{off}}$ is accepted when $\textsc{Coverage}(T_{\mathrm{off}},T'_{\mathrm{off}})$ passes---every designed motif is present and few spurious motifs are introduced. Otherwise, the missing or spurious motifs are returned as feedback, and the realization is retried, up to a budget of five attempts. This is a consistency check, not a measure of quality: it certifies that a candidate is built from the motifs it was designed to contain, not that it is a good magnet compound. Whether a candidate is stable, novel, and magnetic is decided downstream, where the ERS ranks it (Methods~5) and, if selected, first-principles calculations validate it (Methods~6), within the run described in Methods~1.

\subsection*{4. Motif clustering}

We grouped the motifs into a small number of recurring geometric archetypes. Because a motif is text rather than a vector, ordinary distance-based clustering does not apply. Instead, we used an agentic version of $K$-means (Algorithm~\ref{alg:cluster})~\cite{monti2003}, in which each group is represented by a \emph{prototype}---a short paragraph summarizing the motifs it contains, analogous to a centroid. Language-model workers voted on the group to which each motif was best matched (assignment), and the update step resummarized each prototype from its members and adjusted the number of groups. Only the categorical votes and prototype summaries came from the model; all counting and every merge, split, and stopping decision were deterministic. Thus, no numerical judgment rested with the model, and $K$ was discovered rather than fixed.
\emph{Assignment (E-step).} Each motif, shown alongside the current group prototypes, was voted on by $R=5$ independent workers (Claude Fable~5 at temperature $0.5$; five workers trade off robustness against cost), with each worker choosing the best-matching group or \emph{none}. If $n_{ig}$ workers chose group $g$ for motif $i$, its consensus was the majority share,
\begin{equation}
s_i=\frac{1}{R}\max_{g}\, n_{ig}.
\end{equation}
A high $s_i$ value indicated that the workers agreed on motif $i$; repeated \emph{none} votes flagged a geometry not yet covered by any group.

\emph{Update (M-step).} Each group's prototype was first resummarized from the motifs assigned to it---analogous to moving a $K$-means centroid. Thus, the next round of voting used an up-to-date description. The routine then \emph{merged} two groups that were frequently voted together, \emph{split} a group whose members disagreed, and \emph{birthed} a group from motifs repeatedly voted \emph{none}. Two ratios drove these operations---the confusion $c_{gh}$ shared by groups $g$ and $h$, and the contested fraction $\rho_g$ of low-consensus members (consensus below $0.6$) in a group:
\begin{equation}
c_{gh}=\frac{|g\sqcap h|}{\min(|g|,|h|)},\qquad \rho_g=\frac{|\{i\in g:\,s_i<0.6\}|}{|g|},
\end{equation}
where $g\sqcap h$ is the set of motifs voted into both groups and $|g|$ is the size of group $g$. Groups were merged when $c_{gh}\ge0.5$ (over at least three motifs) and split when $\rho_g>0.3$, unless the contested votes were concentrated on one rival group, in which case the two were merged instead. Motifs that workers repeatedly voted \emph{none} matched no current group. Once at least 15 such motifs had accumulated, the model analyzed them and summarized them into one or more new groups, which entered the next round of voting. Thus, $K$ changed only when the votes warranted it. Groups were defined at the archetype level---unit class, coordination geometry, and coordination-number band---so that each represented a recognizable building block.

\emph{Convergence.} The steps alternated until the groups stopped changing, as judged by the mean consensus $\bar{s}=|\mathcal{M}|^{-1}\sum_i s_i$ and the agreement between assignments in successive rounds, measured by the Adjusted Rand Index (ARI)~\cite{hubert1985},
\begin{equation}
\mathrm{ARI}=\frac{\sum_{gh}\binom{n_{gh}}{2}-\big[\sum_g\binom{a_g}{2}\sum_h\binom{b_h}{2}\big]\big/\binom{n}{2}}{\tfrac{1}{2}\big[\sum_g\binom{a_g}{2}+\sum_h\binom{b_h}{2}\big]-\big[\sum_g\binom{a_g}{2}\sum_h\binom{b_h}{2}\big]\big/\binom{n}{2}},
\end{equation}
with $n_{gh}$ denoting the motifs shared by group $g$ of one round and $h$ of the next, $a_g$ and $b_h$ denote their sizes, and $n$ is the total number of motifs. Clustering stopped when $\bar{s}\ge0.8$ and $\mathrm{ARI}\ge0.95$ for two rounds, or after eight rounds.

\begin{algorithm}[tbp]
\caption{\proc{MotifCluster}: agentic consensus clustering}\label{alg:cluster}
\footnotesize
\textbf{Input:} a set of motifs $\mathcal{M}$\\
\textbf{Output:} groups $\mathcal{G}$ ($K$ discovered) and an assignment of every motif in $\mathcal{M}$
\begin{pseudo}
\item $\mathcal{G}\gets$ initial prototypes proposed from $\mathcal{M}$ (approximately a dozen)
\item \kw{repeat}
\item \ind \textbf{E-step.} $5$ workers vote every motif of $\mathcal{M}\to$ a group or \emph{none}; compute consensus $s_i$
\item \ind \textbf{M-step.} resummarize each group's prototype from its members;
\item \indd \kw{merge} groups with confusion $c_{gh}\ge0.5$;
\item \indd \kw{split} a group with contested fraction $\rho_g>0.3$;
\item \indd \kw{birth} new groups from $\ge15$ motifs voted \emph{none}
\item \kw{until} (mean consensus $\bar{s}\ge0.8$ \kw{and} $\mathrm{ARI}\ge0.95$, two rounds) \kw{or} $8$ rounds
\item \kw{return} $\mathcal{G}$ and the converged assignment of $\mathcal{M}$
\end{pseudo}
\end{algorithm}

\subsection*{5. Evidence-based Recommender System (ERS)}

Within each generation, the ERS \emph{ordered} the candidates produced by the agents so that costly first-principles validation was performed first on the most promising candidates. It did not determine stability---that was the role of DFT (Methods~6)---but supplied an inexpensive, uncertainty-aware prior over which candidates were worth relaxing.
The ERS scored each candidate based on its motif-group composition using Dempster--Shafer evidence theory, following the descriptor-free recommender introduced for alloy discovery~\cite{Ha2021,Ha2025}. The 17 motif groups defined in Methods~4 serve as chemical elements in that framework. The groups were fixed after clustering, and each candidate's motifs were assigned to them before scoring, ensuring that the vocabulary over which the ERS ranked candidates remained consistent throughout the run. Because each motif group represented a class of geometrically similar building blocks, substituting one group for another was a physically reasonable operation---exchanging one structural building block for another---and the ERS learned from the observed structures which such substitutions preserved the target property.

Each observed structure was represented by the \emph{set} $S=\{g_1,\dots,g_k\}$ of the motif groups it contained, with each $g_i\in\mathcal{G}$ belonging to the $17$-group vocabulary of Methods~4. Multiplicity, connectivity, composition, and lattice information below this resolution were deliberately not encoded, so the belief returned by the ERS was a prior over motif-set plausibility rather than a thermodynamic prediction. Because composition was among the omitted variables, the group co-occurrences it learned partly reflected shared chemistry rather than structural interchangeability alone. Each structure carried a binary target label $y_S\in\{\top,\lnot\top\}$ with $y_S=\top$ when $S$ met the design objective. For the stability target used to prioritize the run, $y_S=\top$ when $E_\mathrm{hull}(S)\le\varepsilon$ at the same $\varepsilon=0.1$~eV/atom used throughout---a metastability window rather than strict on-hull stability (Supplementary Information~S4.1). The evidence set $\mathcal{D}$ consisted of the labeled seed structures and was augmented each generation by the DFT-validated candidates, thereby allowing the recommender to refresh as the search progressed. Evidence was accumulated in two stages---the substitutability of motif-group combinations, followed by the target belief of a candidate---with each represented by a Dempster--Shafer mass function fused by Dempster's rule.

\emph{Substitutability of motif-group combinations.} Any two observed structures sharing at least one motif group, $S_i\cap S_j\neq\emptyset$, provide one piece of evidence as to whether the group combinations by which they differ are interchangeable. By using $C_t=S_i\setminus S_j$ and $C_v=S_j\setminus S_i$ to refer to the differing combinations against the shared context $S_i\cap S_j$, and taking a frame of discernment $\Omega_{\mathrm{sub}}=\{\mathrm{sub},\lnot\mathrm{sub}\}$, the pair contributes
\begin{align}
m^{C_t,C_v}_{S_i,S_j}(\{\mathrm{sub}\}) &= \alpha\,[\,y_{S_i}=y_{S_j}\,],\\
m^{C_t,C_v}_{S_i,S_j}(\{\lnot\mathrm{sub}\}) &= \alpha\,[\,y_{S_i}\neq y_{S_j}\,],\\
m^{C_t,C_v}_{S_i,S_j}(\{\mathrm{sub},\lnot\mathrm{sub}\}) &= 1-\alpha,
\end{align}
where $[\,\cdot\,]$ is the Iverson bracket and the discount $\alpha\in(0,1)$ leaves a residual mass $1-\alpha$ on the full frame, encoding the uncertainty of a single observation. Combining all pieces of evidence bearing on a given pair of combinations over the data set $\mathcal{D}$ by Dempster's rule (below) gives a substitutability belief $M(C_t,C_v)=m^{C_t,C_v}_{\mathcal{D}}(\{\mathrm{sub}\})$, symmetric in its two arguments.

\emph{Target belief of a candidate.} A candidate $S_{\mathrm{new}}$ is scored over every observed host $S_{\mathrm{host}}\in\mathcal{D}$ and every single substitution $C_t\!\leftarrow\!C_v$, with $C_t\subseteq S_{\mathrm{host}}$, that yields $S_{\mathrm{new}}=(S_{\mathrm{host}}\setminus C_t)\cup C_v$ and whose differing pair carries nonzero substitutability belief. Over the target frame $\Omega=\{\top,\lnot\top\}$, each such host contributes
\begin{align}
m^{S_{\mathrm{new}}}_{S_{\mathrm{host}},C_t\leftarrow C_v}\!(\{\top\}) &= M(C_t,C_v)\,[y_{S_{\mathrm{host}}}{=}\top],\\
m^{S_{\mathrm{new}}}_{S_{\mathrm{host}},C_t\leftarrow C_v}\!(\{\lnot\top\}) &= M(C_t,C_v)\,[y_{S_{\mathrm{host}}}{=}\lnot\top],\\
m^{S_{\mathrm{new}}}_{S_{\mathrm{host}},C_t\leftarrow C_v}\!(\{\top,\lnot\top\}) &= 1-M(C_t,C_v),
\end{align}
a Shafer discount in which the host transfers its own label to the candidate with a strength equal to the substitutability belief, with the residual $1-M(C_t,C_v)$ left as ignorance.

\emph{Combination and ranking.} Pieces of evidence are fused by Dempster's rule of combination,
\begin{equation}
\begin{aligned}
(m_a\oplus m_b)(\omega)&=\frac{1}{1-K}\sum_{\omega_k\cap\omega_h=\omega} m_a(\omega_k)\,m_b(\omega_h),\\[2pt]
K&=\sum_{\omega_k\cap\omega_h=\emptyset} m_a(\omega_k)\,m_b(\omega_h),
\end{aligned}
\end{equation}
which is commutative and associative. Thus, the fusion is independent of the order in which hosts are combined. Fusing all hosts determines a candidate's final mass $m^{S_{\mathrm{new}}}_{\mathcal{D}}$, and candidates are ranked in descending order of the likelihood that they meet the target, $m^{S_{\mathrm{new}}}_{\mathcal{D}}(\{\top\})$. The discount $\alpha$ is fixed by a grid search that maximizes the cross-validated reproduction of the observed labels ($\alpha=0.1$), with the ranking being largely insensitive to it, as in the alloy setting~\cite{Ha2021}.

\subsection*{6. Property evaluation and first-principles validation}

The validated quantities---the relaxed structure, total energy, energy above the convex hull, and target property---are recorded, and the structure is returned to the database.

\textbf{DFT Relaxation.} Final structural optimization is performed using the Vienna Ab initio Simulation Package (VASP, version 5.4.4)~\cite{vasp1,vasp2,vasp3} with spin-polarized calculations (ISPIN = 2) to capture the magnetic nature of the rare-earth--transition-metal compounds studied. The GGA with the Perdew--Burke--Ernzerhof functional~\cite{perdew1996generalized} is employed to describe exchange-correlation effects. Initial magnetic moments are assigned per element: 5.0~$\mu_\mathrm{B}$ for Fe, Mn, Ni, Cr, V, Mo, W and Ce, 10.0~$\mu_\mathrm{B}$ for Eu, and 0.6~$\mu_\mathrm{B}$ for every other element for which a moment was defined (Supplementary Information~S8). The projector augmented-wave (PAW) method~\cite{Blochl94,Kresse99} is employed with Sm\_3 and Fe\_pv potentials to properly describe the valence-electron configurations, using a plane-wave energy cutoff of 520~eV. Nonspherical contributions to the PAW spheres (LASPH = .TRUE.) and extended angular-momentum mixing (LMAXMIX = 6) are enabled for accurate charge-density mixing in these open-core rare-earth systems (the Sm 4$f$ electrons being treated in the core by the Sm\_3 potential, with their moment restored as described below). Brillouin zone sampling employs $\Gamma$-centered $k$-point meshes with KSPACING = 0.25~\r{A}$^{-1}$. These settings follow established high-throughput practice for intermetallic convex-hull construction and match the Materials Project workflow used to compute the reference energies. Each candidate's energy above the convex hull is measured against a consistently computed hull, and the $0.1$~eV/atom stability cutoff has the same meaning for generated candidates as for the reference database. The potentials are sourced from the POTCAR library version 5.4 of VASP~\cite{vasp_pot,Kresse99,Blochl94}. The potential and initial magnetic moment of every element, and the versions of the Materials Project and Alexandria databases used as references, are listed in Supplementary Information~S8.

Electronic self-consistency is achieved with an energy-convergence criterion of EDIFF = $10^{-4}$~eV, while ionic relaxation stops when the total energy changes by less than $10^{-3}$~eV between ionic steps (the VASP default, EDIFFG = $10\times$EDIFF). Full structural relaxation including ionic positions, cell shape, and cell volume (ISIF = 3) is performed using the conjugate gradient algorithm (IBRION = 2) for up to 50 ionic steps (NSW = 50) with a maximum of 100 electronic self-consistency iterations per ionic step (NELM = 100). Electronic structure calculations use the tetrahedron method with Bl\"{o}chl corrections (ISMEAR = $-$5). The FAST algorithm is employed for electronic minimization, and orbital-decomposed density of states are computed (LORBIT = 11) for the postanalysis of magnetic properties.

\textbf{Formation-energy Calculation.} Thermodynamic stability is assessed through formation-energy calculations. The formation energy per atom, $\Delta H_f$, is computed as follows:
\begin{equation}
\Delta H_f = \frac{1}{N}\left(E_{\mathrm{total}} - \sum_i n_i \mu_i\right),
\end{equation}
where $E_{\mathrm{total}}$ is the DFT total energy of the relaxed structure, $N$ is the total number of atoms, $n_i$ is the number of atoms of element $i$, and $\mu_i$ is the chemical potential of element $i$ referenced to its standard elemental ground state. The formation energy serves as a key indicator of the stability of candidate compounds~\cite{Matsumoto20}.

\textbf{Energy Above Hull.} Thermodynamic stability relative to competing phases is quantified by the energy above the convex hull. For each generated structure, a convex hull is constructed for its chemical system from Materials Project entries spanning the constituent elements. The energy above the hull is then evaluated as the difference between the structure's DFT formation energy per atom (relative to elemental references calculated with the same settings) and hull energy at its composition, the latter is taken as the proportion-weighted formation energies of the structure's decomposition products on the hull. A value of zero denotes an on-hull (thermodynamically stable) structure, and positive values denote metastability.

\textbf{Magnetization Calculation.} The total magnetic moment $\mu[x]$ of a structure is the DFT moment corrected for the lanthanides whose 4$f$ electrons are treated in the core (Supplementary Information~S8)~\cite{Miyake18}:
\begin{equation}
\label{eqn:magnetization}
\mu[x] = M[x] + \sum_{k\in\mathrm{Ln_{c}}} \sigma_{k}\, g_{J_{k}} J_{k},
\end{equation}
where $M[x]$ is the total moment of the cell from DFT, which already contains the 4$f$ moments of Ce, Eu, and Gd, whose 4$f$ electrons are in the valence; the sum runs over the lanthanides $\mathrm{Ln_{c}}$ whose 4$f$ electrons are in the core and restores each one's 4$f$ moment analytically: $g_{J_{k}}J_{k}$ is its Hund's-rule ground-state moment ($g_{J_{k}}$ is the Land\'{e} factor and $J_{k}$ is the total angular momentum), and $\sigma_{k}=\pm1$ denotes its coupling to the transition-metal sublattice. The 4$f$--3$d$ exchange holds the rare-earth spin antiparallel to the transition-metal moment. Therefore, by Hund's third rule, the net 4$f$ moment \emph{adds} for the light lanthanides ($\sigma_{k}=+1$; Pr, Nd, Sm, whose 4$f$ shell is less than half-filled) and \emph{opposes} it for the heavier ones ($\sigma_{k}=-1$; Tb, Dy, Ho, Er): $g_{J}J=0.714~\mu_\mathrm{B}$ for Sm$^{3+}$ is added, whereas the larger heavy-lanthanide moments ($\sim\!9$--$10~\mu_\mathrm{B}$) are subtracted; Yb, treated as divalent, has a filled 4$f$ shell and contributes none. We neglect the $J$-mixing enhancement of the Sm$^{3+}$ moment, which is small on the scale of the $1$~T threshold~\cite{Miyake18}. The saturation magnetization is the magnitude of this net moment, $\mu_{0}M_{s}[x] = \mu_{0}\,\lvert\mu[x]\rvert/V$, where $V$ is the unit cell volume and $\mu_{0}$ is the vacuum permeability.


\section*{Author Contributions}

\textbf{D.-K.L.}: Conceptualization, Methodology, Software, Validation, Investigation, Writing -- review \& editing. 
\textbf{M.-Q.H.}: Conceptualization, Methodology, Software, Investigation, Formal analysis, Writing -- original draft. 
\textbf{H.-P.V.-D.}: Investigation, Validation.
\textbf{V.-C.N.}: Software, Validation.
\textbf{H.T.}: Conceptualization, Writing -- review \& editing.
\textbf{T.T.}: Conceptualization, Writing -- review \& editing.
\textbf{T.M.}: Conceptualization, Investigation, Validation, Writing -- review \& editing.
\textbf{H.K.}: Conceptualization, Investigation, Validation, Writing -- review \& editing.
\textbf{H.-C.D.}: Conceptualization, Supervision, Resources, Methodology, Writing -- review \& editing.

\section*{Conflicts of Interest}

The authors declare no competing interests.

\section*{Data Availability}

All data generated and analysed in this study, including the DFT-relaxed structures, are deposited in Zenodo~\cite{matevolve_data}.

\section*{Code Availability}
The code implementing MatEvolve, including the complete prompt templates and the scripts that compute the benchmark rates, is available on GitHub~\cite{matevolve_github} under an MIT license.

\section*{Acknowledgements}

This work is supported by the JST-CREST Program (Innovative Measurement and Analysis), under grant number JPMJCR2235; the MEXT Program (Data Creation and Utilization-Type Material Research and Development Project), under grant number JPMXP1122715503; the JST "Adopting Sustainable Partnerships for Innovative Research Ecosystem (ASPIRE)" Project, under grant number JPMJAP2321; and JSPS KAKENHI grant numbers 20K05301, JP19H05815, 23KJ1035, 23K03950, and JP23H05403.


\bibliographystyle{elsarticle-num-names}
\bibliography{main}

\end{document}


\maketitle
\section{CIF2Text and the motif-profile representation}
\label{si:agents}\label{si:grammar}

\paragraph{\textsc{CIF2Text} prompts.} \textsc{CIF2Text} reads a CIF and, aided by a deterministic geometry tool (neighbour lists within $3.5$~\AA\ under periodic boundary conditions and Voronoi coordination analysis), returns a JSON motif profile comprising a list of motifs (the count is set by the model, typically $5$--$10$) and a hard-magnet assessment. Each motif's raw description is then rewritten into the geometry-only controlled vocabulary described below by a second standardization pass. Only the composition and the motif descriptions are passed onward---to \textsc{Text2CIF} for reconstruction and design realization (SI~S2) and to the design agent (SI~S3), via the \texttt{\{composition\}} and \texttt{\{text\_description\}} fields of their prompts. No numerical lattice parameter, space-group symbol or structure-type label therefore crosses the round trip.

 The system and user prompts are reproduced below (typographic rules and arrows normalized to ASCII); the user prompt's \texttt{\{cif\_content\}} field is filled at run time.

\begin{lstlisting}[style=prompt]
You are an expert materials scientist specializing in crystal structure
analysis and magnetic materials. Our work focus on RE-TM-X systems.

Describe structures the way a human expert does:
- Start from physically meaningful units, not raw coordinates
- Pair structural observation with physical implication immediately
- Use known references (Fe-Fe ~2.5A, Nd2Fe14B benchmarks, etc.)
- Flag anomalies vs. RE-TM-X family norms
- Reason: structure -> interaction -> property in continuous narrative

The text description need to be useful to infer structure-property relationship, and reconstruct CIF.

================================================================================
STRUCTURAL PATTERNS (MOTIFS)
================================================================================

As a materials scientist, identify notable structural patterns that provide
insight into this material's properties.

A motif is a multi-atom structural unit carrying specific function.
Use your expert judgment - not every structure has the same number of motifs.

Patterns to consider (select what's relevant):
- Sublattices: RE/TM spatial arrangement (BCC, FCC,..., geometry, distances, interactions)
- Extended: Layers/chains/rods with specific composition/function
- Structural anchor: Which sublattice forms the global scaffold
- Networks: Connected frameworks (Fe-B, Fe-Fe beyond simple bonds)
- Cages/polyhedra: Coordination environments around central atoms which you believe it be important.
- Clusters: Local structure which contributed with one center atom and its environment (Fe4B, not generic Fe4; ring Fe6; Sm-Fe12 polyhedron)


Skip:
- Single atoms or simple dimers already covered in network description
- Generic motifs common across many materials

Motifs may overlap in atoms - focus on distinct functional roles.
*Note:* (IMPORTANT)
- Motifs may span across unit cell boundaries - consider periodic images
when identifying coordination environments and clusters.
- Number of motifs: Depend on your knowledge and your understanding. It should be from 5-10.
- Describe geometrical information should support the design/reconstruct geometrical structure/CIF

For each motif:
- name: descriptive name of the pattern
- description: 4-6 sentences covering:
  (1) geometry - describe geometrical shape, coordination, bond lengths/angles vs. references, which support enough to reconstruct geometry
  (2) functional role - exchange/anisotropy/stability
  (3) mechanism - direct exchange/RKKY/superexchange/CEF
  (4) property impact - effect on Tc, Ms, or K1

================================================================================
OUTPUT FORMAT (JSON): FOLLOW STRICTLY THE PROVIDED JSON FORMAT
================================================================================

{
  "motifs": [
    {
      "motif_id": "..."
      "name": "short name",
      "description": "..."
    }
  ],
  "hard_magnet_assessment": {
    "anisotropy": "strong/weak/absent - reason",
    "exchange": "FM/AFM/neutral - mechanism",
    "stability": "STABLE/UNSTABLE - basis",
    "verdict": "hard/soft/uncertain - one sentence"
  }
}
\end{lstlisting}
\begin{lstlisting}[style=prompt]
## CIF FILE
{cif_content}

## TASK
1. Compute bond lengths for atom pairs within 3.5A (apply periodic
   boundary conditions), determine local structures (for example: using voronoi)
2. Identify notable structural patterns as motifs - use your expertise
   to determine what's significant for this specific structure
3. Complete hard_magnet_assessment

Return ONLY valid JSON. Do not add additional keys.
\end{lstlisting}

Materials scientists reason about a crystal---and about the origin of its properties---in terms of recurring local structural motifs, such as coordination polyhedra, cages and characteristic bonded clusters, and the way these connect, rather than in terms of raw atomic coordinates. For the properties targeted here---thermodynamic stability and magnetization---it is the presence and arrangement of such motifs that drives the behaviour. We therefore represent a structure by its \textit{motif profile}: the set of distinct motifs it contains, written as a natural-language paragraph together with the chemical composition. Each motif is described by its central species, the identity and geometry of the neighbours that coordinate it, and the length regime of its defining bonds (short, typical, or long), expressed in a controlled vocabulary so that the profiles of different structures are written in common terms and can be compared and edited. The profile records no atomic coordinates, no lattice parameters, no space-group symbol and no structure-type label: it states which building blocks a crystal is made of and how they are bonded, not where each atom sits.

Identifying the chemically meaningful motifs of an arbitrary crystal is itself expert knowledge and is difficult to capture in fixed rules. We delegate it to a large language model that has been trained on a broad corpus of scientific literature and is prompted to act as a materials scientist, reading the structure and naming its motifs. Expressing the result as text---the medium in which the model reasons---is what lets the downstream agents---Motif editing and Motif Translate---operate on the representation directly. The controlled vocabulary and the exact bond-length regimes it uses are defined in the remainder of this section.

\textsc{CIF2Text} names motifs in a geometry-only controlled vocabulary. After the descriptive pass above, a second standardization pass rewrites each motif entry into square-bracketed tags drawn solely from this vocabulary, so that motifs from different structures are written in common terms and can be matched and clustered; the vocabulary carries only shape, coordination, connectivity and periodicity, and deliberately excludes all property, electronic-structure and magnetic terms (those are reserved for the design agent, SI~S3).

Each entry declares exactly one \emph{motif type} (\texttt{polyhedron}, \texttt{cage}, \texttt{cluster}, \texttt{network}, \texttt{layer}, \texttt{chain}, \texttt{ring}, \texttt{dimer}, \texttt{isolated\_site}, \texttt{void}, \texttt{sublattice} or \texttt{interpenetrating}). Each atom carries a geometric \emph{role} (\texttt{center}, \texttt{vertex}, \texttt{edge}, \texttt{face}, \texttt{bridge}, \texttt{terminal}, \texttt{spacer}, \texttt{interstitial} or \texttt{framework}). The \emph{coordination geometry} is named from a fixed set graded by coordination number---low-CN (e.g.\ \texttt{linear}, \texttt{tetrahedral}, \texttt{square-planar}), medium-CN (e.g.\ \texttt{octahedral}, \texttt{square-antiprismatic}, \texttt{tri-capped-trigonal-prismatic}) and high-CN (\texttt{cuboctahedral}, \texttt{icosahedral}, \texttt{Frank-Kasper-Z14/Z15/Z16}), with \texttt{distorted-} and \texttt{mono/bi/tri-capped-} qualifiers---and reported alongside a CN bucket (\texttt{[low CN]} 2--4, \texttt{[medium CN]} 5--8, \texttt{[high CN]} 9--16, \texttt{[very high CN]} $>$16), a shell composition (\texttt{[pure/binary/mixed shell]}), a connectivity tag (\texttt{corner/edge/face/vertex-sharing}, \texttt{isolated}, \texttt{percolating}, \texttt{chained}, \texttt{layered}, \texttt{framework}, \texttt{interpenetrating}), a dimensionality (\texttt{[0D]}--\texttt{[3D]}, with \texttt{quasi-} variants) and, when known, the site symmetry as a Hermann--Mauguin symbol.

Bond distances are made element-pair agnostic by comparing the observed distance $d$ to the sum of covalent radii $r_{\mathrm{sum}}=r_{\mathrm{cov}}(A)+r_{\mathrm{cov}}(B)$ of the bonded pair~\cite{cordero2008} (pymatgen radii may be used as a deterministic substitute); no raw \AA\ value appears in a standardized description. The regimes are given in Table~\ref{tab:bondregime}:

\begin{table}[h]\centering\small
\caption{Bond-distance regimes of the motif grammar.}
\label{tab:bondregime}
\begin{tabular}{lll}
\hline
Tag & Condition & Meaning\\
\hline
\texttt{[short bond]} & $d<0.90\,r_{\mathrm{sum}}$ & compressed, strong overlap\\
\texttt{[typical bond]} & $0.90\,r_{\mathrm{sum}}\le d\le 1.10\,r_{\mathrm{sum}}$ & standard bonding\\
\texttt{[long bond]} & $1.10\,r_{\mathrm{sum}}<d\le 1.30\,r_{\mathrm{sum}}$ & stretched, weak\\
\texttt{[non-bonded]} & $d>1.30\,r_{\mathrm{sum}}$ & no direct bond\\
\hline
\end{tabular}
\end{table}

Every standardized motif entry is one paragraph following a fixed template: sentence~1 declares the motif, its roled atoms, coordination, CN bucket, shell and dimensionality/topology; sentence~2 gives optional spatial or stacking context; sentence~3 begins with a bond-pair label and states its regime and the site symmetry; sentence~4 optionally adds a connectivity note. The standardizer's system and user prompts are reproduced below.

\begin{lstlisting}[style=prompt]
You are a crystal-structure motif standardizer. Convert raw motif descriptions
into a controlled-vocabulary form that captures **3D geometry only** -- shape,
coordination, connectivity, periodicity. Do not introduce any property,
electronic-structure, or magnetic terms.

### Hard rules

- All vocabulary in square brackets comes from `motif_vocabulary.md` -- no other tags.
- Element symbols appear bare immediately after their geometric role label: `[center] Ti`, `[vertex] O`.
- Convert raw bond distances to the regime labels in Section 6 of the vocabulary using the sum of covalent radii of the bonded pair. **Never write a raw A value in the output.**
- Each motif becomes 2-4 sentences. Use one continuous paragraph per motif.
- Only include elements explicitly named in the raw motif text. Do not infer or carry over elements from other motifs.

### Grammar

Every standardized motif follows this structure:

S1: A [MOTIF] of [ROLE] <Element>, [ROLE] <Element>, ... in [COORDINATION]
    coordination with [CN bucket], [SHELL], [DIMENSIONALITY] [TOPOLOGY].
S2: Optional spatial / stacking context -- where the motif sits relative to
    the cell or to neighbouring motifs.
S3: [BOND-PAIR] in [REGIME] regime; site symmetry is [point group].
S4: Optional -- connectivity note ("shares corners with adjacent motifs",
    "isolated from other motifs", etc.) when not already covered in S1.

S3 **must begin** with a bond-pair label, not with the regime alone.

Wrong: `[short bond] regime between Ti and O ...`

Right: `[Ti-O] in [typical bond] regime; site symmetry is [m-3m].`

### Hard rules that are commonly broken

1. **No magnetic / electronic terms.** Words to never use: ferromagnetic, antiferromagnetic, ferrimagnetic, exchange, anisotropy, hybridization, CEF, Stoner, RKKY, f-d, d-d, superexchange, moment, spin-orbit, easy axis, magnetic, magnetization, Curie, Neel.
2. **No element-role assignments by property.** Use only the geometric roles in Section 2 of the vocabulary -- `center`, `vertex`, `edge`, `face`, `bridge`, `terminal`, `spacer`, `interstitial`, `framework`. Never `magnetic TM`, `heavy RE`, etc.
3. **No raw distances in A.** Convert to `[short bond]`, `[typical bond]`, `[long bond]`, `[non-bonded]` using the covalent-radius rule.
4. **One motif type per entry.** S1 declares exactly one motif type.

---
\end{lstlisting}
\begin{lstlisting}[style=prompt]
## TASK
Standardize each motif below using the controlled vocabulary from motif_vocabulary.md.
Return one paragraph per motif in the format:

Motif #<id>: <standardized paragraph>

## MOTIFS

Motif #1: <raw description>
Motif #2: <raw description>
...
\end{lstlisting}
\section{Text2CIF: prompts, validation and iteration budget}
\label{si:generate}
\textsc{Text2CIF} maps a design concept---an integral composition and a text description of the target motifs---to a CIF. That a profile can be turned back into a crystal at all rests on an analogy: a motif profile states a geometric object in words, as a problem in plane or solid geometry is stated and routinely solved without coordinates, so a competent reasoner---human or model---can construct a figure consistent with it, and \textsc{Text2CIF} uses a language-model agent in exactly this role. Because the profile fixes the motifs but not the coordinates, the realization is under-determined and a single profile may admit more than one valid crystal. Each proposed CIF is parsed and checked deterministically for validity: parseable syntax, full site occupancy ($1.0$), integral atom counts, and a minimum interatomic distance above $1.5$~\AA. On failure the specific problem is returned to the agent (marked \texttt{\#\#ISSUE\#\#}) and generation is retried, up to a budget of five attempts; a valid CIF is then pre-relaxed with M3GNet, reduced to its primitive cell, and re-translated for the motif-coverage check of Methods~2. This five-attempt budget is the inner validity loop; Structure Translate (Methods~2, Algorithm~2) and Motif Translate (Methods~3, Algorithm~3) each wrap it in an outer retry loop of the same budget. The system and user prompts are reproduced below; the user prompt's \texttt{\{composition\}} and \texttt{\{text\_description\}} fields are filled at run time.

\begin{lstlisting}[style=prompt]
You are a crystallographic structure generator for RE:TM:X magnetic intermetallic compounds.

ELEMENT DEFINITIONS:
- RE (Rare Earth): Sm, Nd, Ce, Pr, Gd, Tb, Dy, Ho, Er, Y - provide magnetocrystalline anisotropy
- TM (Transition Metal): Fe, Co, Ni (magnetic); Ti, V, Zr, Mo, W (stabilizing)
- X (Interstitial): N, C, B, H - occupy interstitial sites
- M (Metalloid): Si, Al, Ga, Ge, Sn - substitute at TM sites for stability

INPUT INFORMATION: one design concept contains:
   - Composition: Target integral formula
   - Text description: Text description with the motifs of the ideal structure
YOUR APPROACH:

STEP 1: PARSE THE DESIGN CONCEPT
  - Identify target composition
  - List all motifs and their origins:
STEP 2: BUILD THE NEW STRUCTURE (GENERATE CIF)
  - Choose appropriate space group based on motif symmetries
  - Include all crystallographic data
  - Full occupancy (1.0) at all sites
  - Integral atom counts only

STEP 3: VALIDATE MOTIF REALIZATION
  - Verify each designed motif is present in output structure
  - Check coordination geometries match motif specifications
  - Ensure no atomic overlaps (min distance > 1.5 A)
  - Confirm composition matches target formula

OUTPUT: Return ONLY the CIF file contents. No explanations.

If you receive ##ISSUE## feedback, analyze the problem and regenerate.
\end{lstlisting}
\begin{lstlisting}[style=prompt]
##DESIGN CONCEPT##:

Composition: {composition}

Text description: {text_description}
\end{lstlisting}
\section{Evolutionary operator over motif profiles}
\label{si:operator}

New candidates are produced by editing motif profiles. The operator (\textsc{EvoOperator}, Methods~3) takes the profiles of one or two parent structures and generates an offspring profile $T_{\mathrm{off}}$ through one of three moves: crossover, combining motifs drawn from two parents; element mutation, substituting the species within a parent's motifs; or geometry mutation, altering a motif's coordination or connectivity. This is the only component that reasons about properties, and it does so only to \emph{propose} edits---the ERS (Methods~5) later ranks the resulting candidates statistically---while the motif profiles that the translation agents pass onward are strictly geometry-only (Section~S1). The resulting offspring profile is emitted as a target for Text2CIF.
The population is initialized from $\mathcal{D}_{\mathrm{seed}}$ and advanced for four generations. In the reported run only the \emph{crossover} operator is used: each design step selects two parents from the current pool---one drawn from the $200$ most stable structures and one from the $200$ of highest saturation magnetization---and produces a single offspring, so that every candidate combines a stability-favoured and a magnetization-favoured parent; the element- and geometry-mutation operators are provided by the framework but were not applied here. Design steps are repeated to populate each generation; the ERS ranks the offspring, and every DFT-validated candidate (Methods~5--6) joins the pool from which the next generation's parents are drawn.

\paragraph{Design-agent prompts.} The design agent returns, as strict JSON, one offspring design---its motif set (\texttt{motifs} in the JSON), an integral composition, and a self-contained text description that \textsc{Text2CIF} can realize without parent context. The system prompt and the two task templates (crossover, which receives two parents; mutation, which receives one) are reproduced below; \texttt{\{operator\}} and \texttt{\{description\_*\}} are filled at run time.

\begin{lstlisting}[style=prompt]
You are a materials design assistant for discovering novel RE:TM:X magnetic intermetallic compounds.

PRIMARY OBJECTIVE:
Design novel magnetic material structures that achieve:
  1. HIGH STABILITY - Thermodynamically stable, resistant to decomposition
  2. SYNTHESIZABILITY - Can be practically synthesized in laboratory
  3. HIGH MAGNETIZATION - Strong magnetic properties (high Ms, high Tc, good anisotropy)

All design decisions must be justified in terms of these three objectives.

INPUT:
For each parent structure, you will receive TEXT REPRESENTATION containing:

   - ### MOTIFS: Local structural patterns with functional descriptions

DESIGN OPERATORS:

You have THREE operators to design improved structures.

===========================================================================
1) CROSSOVER - Combine motifs from multiple parents
===========================================================================

   PURPOSE: Create superior motif sets by selecting from parent motif pools.

   GUIDANCE:
   - Selecting SUBSETS of motifs (not all motifs need to be in every skeleton); REORGANIZING motif combinations (different structural arrangements); MODIFYING motif connections (different attachment patterns)
   - Ensure selected motifs have compatible coordination environments
   - Balance functional roles: exchange, anisotropy, moment, stability
   - Explore different motif combinations to create diverse stoichiometries

===========================================================================
2) ELEMENT MUTATION - Modify elemental composition
===========================================================================

   PURPOSE: Improve properties by changing elements in motifs.

   GUIDANCE:
   - Substitute elements to enhance stability and/or magnetization
   - Add or remove elements to optimize the objectives
   - Slight geometry adjustment allowed if needed

===========================================================================
3) GEOMETRY MUTATION - Modify coordination geometry
===========================================================================

   PURPOSE: Improve properties by changing local geometry of motifs.

   GUIDANCE:
   - Change geometry to enhance functional contributions
   - Modify coordination to improve stability and/or magnetization
   - Slight element adjustment allowed if needed

DESIGN PROCESS:

STEP 1: ANALYZE PARENT MOTIFS
  - For each motif, identify its functional role (exchange, anisotropy, moment, stability)
  - Evaluate how each motif contributes to or limits the objectives
  - Determine which motifs to preserve, modify, or replace

STEP 2: DESIGN MOTIF SET AND COMPOSITION
  - Apply operators (CROSSOVER, ELEMENT MUTATION, GEOMETRY MUTATION)
  - Ensure designed motifs can coexist in a single structure
  - Derive target composition from the designed motif set
  - Write comprehensive reasoning that guides the generator:
    * How motifs were selected/modified from parent(s)
    * Key structural requirements (coordination, sites, distances)
    * Why this design achieves the objectives

CONSTRAINTS:
- Composition MUST be integral (no fractional occupancies)
- NO micro-tuning (no "slight adjustments", "strain", "pressure")
- All motifs must be physically reasonable and coexist in a single structure

OUTPUT FORMAT (STRICT JSON ONLY):
{
  "designs": [
    {
      "motifs": [
         {"motif #<ID of motif. The ID strictly starts from 1>": "<Text description of the motif>"}
      ],
      "composition": "<INTEGRAL formula>",
      "text_description": "Text description of the ideal structure for the idea design"
    }
  ],
}


RULES:
- Text description must be detailed enough for generator to create the correct structure
- NO code fences. Return JSON only.
- *Strict note*: motifs and text_description must be *self-contained* for a generator with no parent context. Do not define motifs by parent indices (e.g., A#2, B#1).
- All motifs must be physically reasonable and coexist in a single structure that improve STABILITY and/or MAGNETIZATION over the parent.
\end{lstlisting}
\begin{lstlisting}[style=prompt]
## TASK:
{operator}

Design one novel structure that improve STABILITY and/or MAGNETIZATION over both parents.


OUTPUT: Return STRICT JSON.
---INPUT PARENTS---

## PARENT A: {description_1}
## PARENT B: {description_2}
\end{lstlisting}
\begin{lstlisting}[style=prompt]
## TASK:
{operator}

Design one novel structure that improve STABILITY and/or MAGNETIZATION over the parent.


OUTPUT: Return STRICT JSON.
---INPUT PARENTS---

## PARENT A: {description_1}
\end{lstlisting}
\section{Evaluation metrics}
\label{si:metrics}

\subsection{Benchmark rates}
\label{si:rates}

All generation methods are scored on a common footing. For a method producing a set $\mathcal{C}$ of DFT-relaxed candidate structures ($N=|\mathcal{C}|=400$ for every method), every rate below is a fraction of $N$. In the head-to-head benchmark each method's 400 DFT relaxations are spent on candidates selected by its own ranking procedure from a disclosed raw-proposal pool (baselines: top 400 of 1{,}000 by CHGNet-estimated formation energy and magnetic moment; MatEvolve: top 100 of 200 per generation over four generations by the ERS), so the reported rates are conditional on each method's selection step---each method evaluated in its intended mode of use. The full proposal pool comprises all 800 MatEvolve proposals, selected and unselected, every one relaxed with DFT.

We fix a stability threshold $\varepsilon=0.1$~eV/atom, a design-target magnetization threshold (for this study) $\tau=1$~T, and a reference database $\mathcal{D}_{\mathrm{ref}}=\text{Materials Project}\cup\text{Alexandria}$. For a candidate $c$ we define the predicates: $\mathrm{stable}(c)$ iff $E_{\text{hull}}(c)\le\varepsilon$; $\mathrm{unique}(c)$ iff $c$ is not a \texttt{StructureMatcher} duplicate of another candidate in $\mathcal{C}$ (one representative per duplicate class is counted); $\mathrm{novel}(c)$ iff $c$ matches no entry of $\mathcal{D}_{\mathrm{ref}}$ under \texttt{StructureMatcher}; $\mathcal{P}(c)$ iff $c$ satisfies the design-target property criterion (here $\mathcal{P}:\mu_0 M_s(c)>\tau$); and $\mathrm{newproto}(c)$ iff $c$ realises a structural prototype absent from $\mathcal{D}_{\mathrm{ref}}$ (defined below).

Following the crystal-structure generation literature~\cite{xie2022cdvae,mattergen2025}, a candidate is \textbf{\SUN} when it is simultaneously \emph{stable}, \emph{unique} and \emph{novel}. On this established base, the design target---a rare-earth-lean permanent-magnet compound beyond the known structure types---sets two further requirements, which we assess separately and together:
\begin{align}
\text{\SUN} &= \bigl\{c\in\mathcal{C} : \mathrm{stable}(c)\wedge\mathrm{unique}(c)\wedge\mathrm{novel}(c)\bigr\},\\
\text{magnetic} &= \bigl\{c\in\text{\SUN}\ :\ \mathcal{P}(c)\bigr\},\\
\text{new prototype} &= \bigl\{c\in\text{\SUN}\ :\ \mathrm{newproto}(c)\bigr\},\\
\text{joint} &{}= \bigl\{c\in\text{\SUN}\ :\ \mathrm{newproto}(c)\wedge\mathcal{P}(c)\bigr\}.
\end{align}
The magnetic and new-prototype sets are subsets of \SUN, but neither contains the other: a structure may be magnetic only, a new prototype only, both, or neither; the joint set is their intersection. The four \emph{rates} reported throughout are the corresponding fractions $|\text{\SUN}|/N$, $|\text{magnetic}|/N$, $|\text{new prototype}|/N$ and $|\text{joint}|/N$: the base \SUN\ rate asks whether a method finds stable new materials at all; the \emph{magnetic} rate asks whether they exhibit the target property; the \emph{new-prototype} rate asks whether they are new arrangements rather than substitutions on known lattices, the requirement for extending the known magnet-compound families; and the \emph{joint} rate asks whether a single candidate meets both requirements, as the design target demands. For the permanent-magnet target of this study, the design-target property is $\mathcal{P}:\mu_0 M_s>1$~T; the $1$~T threshold marks an appreciable ferromagnetic moment of permanent-magnet relevance ($\text{Nd}_2\text{Fe}_{14}\text{B}$ reaches $\mu_0 M_s\approx1.6$~T).

The prototype predicate uses a label $\Lambda$ computed in two steps, identically for candidates and for every reference entry. (i) The structure is symmetrized (pymatgen \texttt{SpacegroupAnalyzer}, \texttt{symprec}~=~0.1~\AA). (ii) It is encoded as an AFLOW prototype string~\cite{Mehl2017,Hicks2021}: anonymized composition, Pearson symbol, space-group number and per-species Wyckoff positions, with species lettered by ascending stoichiometry (ties broken by Wyckoff signature, never element identity). Because species are anonymized, an elemental substitution that preserves stoichiometry, symmetry and site occupation shares its host's label (SmFe$_{12}$ and DyFe$_{12}$ share a label), whereas a change in any of the three yields a new one. Then $\mathrm{newproto}(c)$ holds iff $\Lambda(c)\notin\{\Lambda(r):r\in\mathcal{D}_{\mathrm{ref}}\}$. This single rule is applied to every method and chemistry, so the cross-method comparison is not biased by a magnet-specific taxonomy. The same reference $\mathcal{D}_{\mathrm{ref}}$ and prototype rule apply to the full proposal pool (all 800) and to the benchmark set (the 400 selected by the ERS), so the two are directly comparable.

\subsection{Reconstruction fidelity measures}
\label{si:fidelity}

Experiment~1 (main text, Section~2.3) compares each reconstruction returned by \textsc{Text2CIF} with the original Materials Project structure from which its profile was written. The comparison uses each reconstruction as Motif Translate emitted it, without any DFT relaxation. Four measures are used.

\emph{Match rate.} A reconstruction is counted as matching its original when the pair is accepted by pymatgen's \texttt{StructureMatcher} at the tolerances under which generative models report match rates: \texttt{ltol}~=~0.3, \texttt{stol}~=~0.5 and \texttt{angle\_tol}~=~$10^\circ$.

\emph{Root-mean-square displacement (RMSD).} For every pair, the RMSD is the root-mean-square distance, in \AA ngstr\"oms, between corresponding atomic positions after optimal superposition of the two structures; it is computed for all $948$ pairs.

\emph{Continuous shape measure (CShM).} Shape fidelity is quantified with the continuous shape measure~\cite{pinsky1998cshm,zabrodsky1992csm},
\begin{equation}
\mathcal{S}=100\,\min\frac{\sum_i|q_i-p_i|^2}{\sum_i|q_i-q_0|^2},
\end{equation}
where $q_i$ are the atomic positions of the reference (the original structure), $p_i$ those of the reconstruction, $q_0$ the reference centroid, and the minimization runs over translation, rotation, uniform scaling and atom correspondence. The measure is dimensionless and scores the deviation of a structure from its reference on a scale of $0$--$100$: $\mathcal{S}=0$ denotes an identical match and larger values greater distortion. Values were computed with SHAPE and cosymlib~\cite{llunell2013shape,carreras2024cosymlib}.

Two remarks govern how the CShM is read in the main text. First, because the denominator is the second moment of the point set about its centroid, a given atomic displacement yields a smaller value in a more extended point set; a unit cell is more extended than a single coordination polyhedron, so cell-level values are systematically smaller than polyhedron-level ones for the same displacement. Second, the interpretive bands quoted in Section~2.3 (essentially ideal below about $0.1$, slightly distorted up to $3$, notably distorted from $3$ to $7.5$, and beyond $7.5$ a breakdown of the target geometry~\cite{pinsky1998cshm,alvarez2005shapemaps,shao2012alpo,funes2017butterfly}) were calibrated on single coordination polyhedra. They are therefore used only to order the reconstructions and describe the distribution, not as an acceptance criterion; acceptance within the pipeline is decided by the round-trip tests of Methods~2.

\emph{Local-environment fidelity (OFM).} The orbital-field matrix (OFM)~\cite{Pham2017,Pham2020} encodes the coordination of every atom in a structure---the valence-orbital types of each atom and of its neighbors, weighted by the solid angles of its Voronoi polyhedron---and averages these local descriptors into one fixed-length vector per structure. Because the descriptor is an average over sites, it needs no atom correspondence and applies to all $948$ original--reconstruction pairs. The distance between two structures is the $L_1$ norm of the difference of their OFM vectors. Two quantities are reported in the main text (Section~2.3, Fig.~3c): the distance between each original and its own reconstruction, compared with the distances from that original to its $100$ nearest neighbors among the other originals of $\mathcal{D}_{\mathrm{seed}}$; and, for each original, the rank of its own reconstruction among all other structures (the remaining originals and every reconstruction), from which the fraction of originals whose reconstruction is the nearest neighbor, or lies among the $10$ nearest, is obtained.

\section{Structure map by motif-group substitutability}
\label{si:structuremap}

To relate structures through their motif content and to quantify combinatorial coverage (Fig.~6 of the main text), each structure is represented by the \emph{set} of the 17 motif groups it contains. The distance between two structures $S_i$ and $S_j$ combines their set overlap with the substitutability of the groups by which they differ, following the construction we introduced for alloy compositions~\cite{Ha2021,Pham2020}: $\overline{M}[i,j]=d_{\mathrm{sub}}(S_i,S_j)\,\bigl(1-J(S_i,S_j)\bigr)$, where $J$ is the Jaccard overlap of the two group sets~\cite{jaccard1912} and $d_{\mathrm{sub}}$ is a substitutability distance between the differing groups---small when those groups are readily interchangeable---derived from the Dempster--Shafer substitutability belief $M$ of Methods~5, whose evidence is the labelled seed structures augmented each generation by the DFT-validated candidates, discounted by the same uncertainty parameter $\alpha=0.1$ used there. Structures are embedded from $\overline{M}$ using t-SNE~\cite{tsne}. Combination coverage (Fig.~6b of the main text) is obtained by enumerating, for each order $k\in\{2,3,4\}$, the $\binom{17}{k}$ possible group combinations and labelling each as observed in the seed set, the generated set, both, or neither. Pair stability rates (Fig.~6c of the main text) are the fraction of structures containing a given group pair that lie within $0.1$~eV/atom of the convex hull; pairs whose rate is at least $0.75$ are highlighted. Because all $\binom{17}{2}$ pairs are examined at varying and sometimes small support, this threshold identifies a robustly supported class of compatible pairings rather than individually significance-tested pairs.
\section{Motif-group word clouds}
\label{si:wordcloud}
Figure~5a of the main text labels each of the 17 motif groups by a word cloud of the controlled-vocabulary tags (SI~S1) that recur among its motifs; because the terms are the same bracketed geometry tags that name the motifs, each word cloud reads in the grammar of the representation rather than in free text. The construction (Algorithm~\ref{alg:wordcloud}) has three stages. First, five independent agents extract, from each group's standardized motif descriptions, the vocabulary tags present---reusing the standardizer definitions of SI~S1. Second, the five tag sets of a group are pooled and deduplicated, each tag normalized to its canonical vocabulary form. Third, the pooled tags are weighted by term frequency--inverse document frequency (TF--IDF)---treating a group's tags as one document and the 17 groups as the corpus, so that a tag's weight rises with its frequency within a group and falls with the number of groups sharing it---and the $50$ highest-weighted tags of each group are rendered with type size proportional to TF--IDF weight.

\begin{algorithm}[tbp]
\caption{Motif-group word-cloud construction}\label{alg:wordcloud}
\footnotesize
\textbf{Input:} motif groups $\mathcal{G}$ ($|\mathcal{G}|=17$); standardized motif descriptions (SI~S1); controlled vocabulary $\mathcal{V}$; workers $R=5$\\
\textbf{Output:} the 50 highest-weighted vocabulary tags per group
\begin{pseudo}
\item \kw{for} each group $g\in\mathcal{G}$ \kw{do}
\item \ind \kw{for} $r=1$ \kw{to} $R$ \kw{do}
\item \indd $K_{g,r}\gets\proc{ExtractTags}(\text{descriptions of }g,\ \mathcal{V})$ \cmt{independent agent; SI~S1 standardizer}
\item \ind \kw{end for}
\item \ind $K_g\gets\proc{Dedup}\big(\textstyle\bigcup_r K_{g,r}\big)$ \cmt{pool the 5 agents; normalize to canonical $\mathcal{V}$}
\item \kw{end for}
\item \kw{for} each tag $t$ \kw{do} $\mathrm{df}(t)\gets|\{g:\,t\in K_g\}|$ \cmt{groups containing $t$}
\item \kw{for} each group $g$ and each tag $t\in K_g$ \kw{do}
\item \ind $\mathrm{tf}(t,g)\gets$ occurrences of $t$ among group $g$'s motif tags
\item \ind $w(t,g)\gets \mathrm{tf}(t,g)\cdot\log\!\big(|\mathcal{G}|/\mathrm{df}(t)\big)$ \cmt{TF--IDF weight}
\item \kw{end for}
\item render the 50 highest-$w(\cdot,g)$ tags of each group $g$, type size $\propto w$
\end{pseudo}
\end{algorithm}

\section{Relation to text-based crystal representations}
\label{si:relatedtext}

Table~\ref{tab:si_textrep} situates the motif profile among prior text-based treatments of crystals, ordered by the direction of the crystal--text translation. Both endpoint directions are established: automated description turns a crystal into text, and text-conditioned generation turns a specification back into a crystal. The motif profile differs from both. Against description, it is written in a geometry-only controlled vocabulary that withholds the coordinates, lattice parameters, space group, and structure-type label that a description reports, and it is certified generative---accepted only if a matching structure can be rebuilt from it (Methods~2). Against text-conditioned generation, its realization step starts not from a composition-and-symmetry prompt but from the detailed motif profile itself, and is one leg of a design loop rather than the goal.

\begin{table}[htbp]
\centering
\caption{Prior text-based treatments of crystals and their relation to MatEvolve.}
\label{tab:si_textrep}
\small
\begin{tabular}{p{3.6cm}p{2.5cm}p{2.9cm}p{5.0cm}}
\hline
\textbf{Work} & \textbf{Direction} & \textbf{Purpose} & \textbf{Relation to MatEvolve} \\
\hline
Robocrystallographer~\cite{ganose2019robocrystallographer} & Crystal $\rightarrow$ text & Structure description & Direct precedent of CIF2Text; its text reports space group and prototype and is an end product, not an editable object \\
LLM-Prop~\cite{rubungo2025llmprop} & Text $\rightarrow$ property & Property prediction & Establishes text as a machine-readable crystal representation \\
Lattice Lingo~\cite{munjal2024latticelingo} & Crystal $\rightarrow$ text $\rightarrow$ property & Multimodal property prediction & Establishes the value of semiglobal textual information \\
Ozawa et al.~\cite{ozawa2026scaledependent} & Representation $\rightarrow$ property & Input-representation comparison & Establishes that representation design governs model performance \\
CrysText~\cite{mohanty2026crystext} & Text $\rightarrow$ crystal & Crystal generation & Direct analogue of Text2CIF, conditioned on composition and symmetry rather than a motif profile \\
\hline
\end{tabular}
\end{table}

Two of these results bear directly on the profile's design. First, Lattice Lingo reports that semiglobal textual information---connectivity and structural arrangement---contributes most to property prediction beyond graph-encoded local structure~\cite{munjal2024latticelingo}. The motif vocabulary (Section~S1) deliberately spans this range, from local coordination units through cages, nets, layers, and chains to connectivity and dimensionality, so the finding supports the vocabulary's reach as a design choice rather than an assertion. Second, Ozawa et al.\ report that input representations carrying space-group information outperform composition-only inputs for property prediction~\cite{ozawa2026scaledependent}; the motif profile nonetheless withholds the space group, deliberately. A space-group label would be invalidated by almost any motif edit, making the profile brittle as an editing medium, and stating it would hand the round-trip test part of its answer, letting the generator reconstruct the lattice without using the motifs. The cost of this omission is what Experiment~1 measures.

\section{First-principles settings and reference databases}

\label{si:dft}
All relaxations used one set of VASP input parameters: PREC = Accurate, ENCUT = 520~eV, EDIFF = $10^{-4}$~eV, ALGO = Fast, NELM = 100, ISPIN = 2, IBRION = 2, ISIF = 3, NSW = 50, EDIFFG at its default ($10^{-3}$~eV, energy-based), ISMEAR = $-5$, KSPACING = 0.25~\AA$^{-1}$ with $\Gamma$-centered meshes (KGAMMA = .TRUE.), LASPH = .TRUE., LMAXMIX = 6, LREAL = Auto and LORBIT = 11. Table~\ref{tab:si_dft} lists the PAW potential and the initial magnetic moment $m_0$ assigned to each element. With the \_3 potentials (Pr, Nd, Pm, Sm, Tb, Dy, Ho, Er, Tm, Lu) the 4$f$ electrons are kept in the core; Ce, Eu and Gd are treated with their standard potentials and Yb with Yb\_2.

\begin{table}[H]
\centering
\caption{PAW potentials (POTCAR library version 5.4) and initial magnetic moments $m_0$ ($\mu_\mathrm{B}$) used in every spin-polarized relaxation; ``--'': no initial moment defined.}
\label{tab:si_dft}
\footnotesize
\setlength{\tabcolsep}{3.5pt}
\begin{tabular}{lll|lll|lll|lll}
\hline
\textbf{El.} & \textbf{PAW} & $\boldsymbol{m_0}$ & \textbf{El.} & \textbf{PAW} & $\boldsymbol{m_0}$ & \textbf{El.} & \textbf{PAW} & $\boldsymbol{m_0}$ & \textbf{El.} & \textbf{PAW} & $\boldsymbol{m_0}$ \\
\hline
Ac & Ac & -- & F & F & -- & Nb & Nb\_pv & 0.6 & Si & Si & 0.6 \\
Ag & Ag & 0.6 & Fe & Fe\_pv & 5.0 & Nd & Nd\_3 & 0.6 & Sm & Sm\_3 & 0.6 \\
Al & Al & 0.6 & Ga & Ga\_d & 0.6 & Ne & Ne & -- & Sn & Sn\_d & 0.6 \\
Ar & Ar & -- & Gd & Gd & 0.6 & Ni & Ni\_pv & 5.0 & Sr & Sr\_sv & 0.6 \\
As & As & 0.6 & Ge & Ge\_d & 0.6 & Np & Np & -- & Ta & Ta\_pv & 0.6 \\
Au & Au & 0.6 & H & H & 0.6 & O & O & -- & Tb & Tb\_3 & 0.6 \\
B & B & 0.6 & He & He & -- & Os & Os\_pv & 0.6 & Tc & Tc\_pv & -- \\
Ba & Ba\_sv & 0.6 & Hf & Hf\_pv & 0.6 & P & P & 0.6 & Te & Te & 0.6 \\
Be & Be\_sv & 0.6 & Hg & Hg & 0.6 & Pa & Pa & -- & Th & Th & 0.6 \\
Bi & Bi & 0.6 & Ho & Ho\_3 & 0.6 & Pb & Pb\_d & 0.6 & Ti & Ti\_pv & 0.6 \\
Br & Br & 0.6 & I & I & 0.6 & Pd & Pd & 0.6 & Tl & Tl\_d & 0.6 \\
C & C & 0.6 & In & In\_d & 0.6 & Pm & Pm\_3 & -- & Tm & Tm\_3 & 0.6 \\
Ca & Ca\_sv & 0.6 & Ir & Ir & 0.6 & Pr & Pr\_3 & 0.6 & U & U & 0.6 \\
Cd & Cd & 0.6 & K & K\_sv & 0.6 & Pt & Pt & 0.6 & V & V\_pv & 5.0 \\
Ce & Ce & 5.0 & Kr & Kr & -- & Pu & Pu & 0.6 & W & W\_sv & 5.0 \\
Cl & Cl & 0.6 & La & La & 0.6 & Rb & Rb\_sv & 0.6 & Xe & Xe & -- \\
Co & Co & 0.6 & Li & Li\_sv & 0.6 & Re & Re\_pv & 0.6 & Y & Y\_sv & 0.6 \\
Cr & Cr\_pv & 5.0 & Lu & Lu\_3 & 0.6 & Rh & Rh\_pv & 0.6 & Yb & Yb\_2 & 0.6 \\
Cs & Cs\_sv & 0.6 & Mg & Mg\_pv & 0.6 & Ru & Ru\_pv & 0.6 & Zn & Zn & 0.6 \\
Cu & Cu\_pv & 0.6 & Mn & Mn\_pv & 5.0 & S & S & 0.6 & Zr & Zr\_sv & 0.6 \\
Dy & Dy\_3 & 0.6 & Mo & Mo\_pv & 5.0 & Sb & Sb & 0.6 & & & \\
Er & Er\_3 & 0.6 & N & N & 0.6 & Sc & Sc\_sv & 0.6 & & & \\
Eu & Eu & 10.0 & Na & Na\_pv & 0.6 & Se & Se & 0.6 & & & \\
\hline
\end{tabular}
\end{table}

\emph{Reference databases.} Novelty and prototype status are assessed against $\mathcal{D}_{\mathrm{ref}}$ = Materials Project (database version v2026.04.13) $\cup$ Alexandria (PBE release 2025.07.02), and the convex hulls are built from the same Materials Project version.

\bibliographystyle{elsarticle-num-names}
\bibliography{main}